\documentstyle[jeep,graphics,equations,twocolumn,doublespace]{article}

\begin{document}           

\newpage

\begin{center}

{\Large {\bf A dynamical theory of speciation on holey adaptive landscapes}}\\ 

\vspace{.25in}

{\bf Sergey Gavrilets}\\

 Departments of Ecology and Evolutionary Biology\\
and Mathematics,\\
University of Tennessee,\\
Knoxville, TN 37996-1610, USA\\ 
phone: (423) 974-8136\\ 
fax:   (423) 974-3067\\
e-mail: gavrila@tiem.utk.edu
\\

\end{center}

key words:   evolution, speciation, holey adaptive landscapes, mathematical
models\\

running head: Dynamical Theory of Speciation\\


ABSTRACT: The metaphor of holey adaptive landscapes provides a 
pictorial representation of the process of speciation as a consequence 
of genetic divergence. In this metaphor, biological populations diverge 
along connected clusters of well-fit genotypes in a multidimensional adaptive 
landscape and become reproductively isolated species when they come to be 
on opposite sides of a ``hole'' in the adaptive landscape. No crossing of 
any adaptive valleys is required. 
I formulate and study a series of simple models describing the dynamics of 
speciation on holey adaptive landscapes driven by mutation and random
genetic drift. Unlike most previous models that concentrate only on some 
stages of speciation,
the models studied here describe the complete process of speciation from initiation until completion. The evolutionary factors included
are selection (reproductive isolation), random genetic
drift, mutation, recombination, and migration. In these models, pre- and 
post-mating reproductive isolation is a consequence of cumulative genetic 
change. I study possibilities for speciation according to allopatric,
parapatric, peripatric and vicariance scenarios. The analytic theory 
satisfactorily matches results of individual-based simulations reported
by Gavrilets et al. (1998). 
It is demonstrated that rapid speciation including simultaneous emergence
of several new species is a plausible outcome of the evolutionary 
dynamics of subdivided populations. 
I consider effects of population size, population 
subdivision, and local adaptation on the dynamics of speciation. I briefly
discuss some implications of the dynamics on holey adaptive landscapes
for molecular evolution.\\

Speciation has traditionally been considered to be  one of the
most important and intriguing processes of evolution. 
In spite of this consensus and significant advances in both experimental
and theoretical studies of evolution, understanding speciation still 
remains a major challenge (Mayr 1982a; Coyne 1992).
The main reason for such a discouraging situation is that direct experimental approaches, which are widely used for solving other problems of evolutionary biology, are not effective for studying speciation because of the time scale 
involved. Experimental work necessarily concentrates on distinct parts of 
the process of speciation intensifying and simplifying the factors under 
study (Rice and Hostert 1993; Templeton 1996). In situations where direct 
experimental studies are difficult or impossible, mathematical modeling 
has proved to be indispensable for providing a unifying framework. 
Although numerous attempts to model parts of the process of speciation 
have 
been made, a quantitative theory of the dynamics of speciation is still 
missing. Currently, verbal theories of speciation are far more advanced 
than mathematical foundations. As often is the case with verbal theories 
(both scientific and otherwise), different deduced (or induced)
aspects of speciation are emphasized by different workers resulting in 
confusion and controversy. The situation is not helped by the absence of 
general agreement on a species definition (e.g., Claridge et al. 1997).

Here, I attempt to develop some foundations for a general dynamical theory of 
speciation. One possible approach to this goal would be to begin with a 
species definition, then to define speciation accordingly and to develop an appropriate dynamical model. I do not think such approach would be very useful,
due to a lack of generality. My models are not based on 
a specific ``species concept``. I reason that  
species are ``different'' with respect to some characteristics, and that
whatever these differences, they have a genetic basis. Thus, modeling 
the dynamics of speciation is equivalent to modeling the dynamics of 
genetic divergence.
I use a bottom-up approach: begin with a model incorporating a range of 
factors thought to lead to speciation (e.g. selection,
mutation, population subdivision etc.) and then try to interpret its 
dynamic behavior in
terms of different ``species concepts''. As expected, many aspects of 
speciation that are emphasized by
different species concepts (such as reproductive isolation, separate
genotypic clusters or common evolutionary trajectories) emerge from the 
same processes. This clearly indicates that 
different species concepts are not mutually exclusive.

The choice of a modeling approach depends upon the purpose of the model. 
A common view in (evolutionary) biology is that
mathematical models are mainly useful for making predictions that can be used
in experimental work. Although such a pragmatic approach is probably
what should be expected in contemporary society, a model's testable
predictions
are not necessarily its main contribution to science. Insights provided
by models, their ability to train our intuition about complex phenomena,
to provide a framework for studying such phenomena and to identify 
key components in complex systems are at least as important as 
specific predictions. For these purposes the most useful tools are 
{\bf simple} models and metaphors.

Sewall Wright's (1932) metaphor of ``rugged adaptive landscapes'' is 
a well-known and widely used metaphor in evolutionary biology. In the standard 
interpretation, a rugged adaptive landscape
is a surface in a multidimensional space that represents the mean fitness
of the population as a function of gamete (or allele) frequencies which
characterize the population state (see Fig.1a). It is envisioned that
this surface has many peaks and valleys corresponding to different 
adaptive and maladaptive population
states, respectively. The population is imagined as a point on the surface 
which
is driven by selection up hill but can get stuck on a local peak.
Two general points about scientific metaphors should be kept in mind.
The first is that specific metaphors (as well as mathematical
models) are good for 
specific purposes only. The second is that accepting a specific metaphor necessarily influences and defines the questions that are
considered to be important. The metaphor of ``rugged adaptive landscapes''
is very useful for thinking about adaptation. However,
its utility for understanding speciation is questionable. From a pragmatic
point of view, the process of splitting a population into two different 
species is impossible to describe in a framework where a population is
the smallest unit. Finer resolution is necessary for describing the
splitting of populations. 
Accepting the metaphor of rugged adaptive landscapes immediately leads to a problem to be solved: how can a population
evolve from one adaptive peak to another across an adaptive valley when
selection opposes any changes away from the current adaptive peak? 
Wright's solution to this problem, his shifting-balance theory (Wright,
1931, 1982), does not seem to be satisfactory (Gavrilets 1996; Coyne et
al. 1997). Provine (1986), Barton and Rouhani (1987), Whitlock et al. 
(1995), Gavrilets (1997a) and Coyne et al. (1997) discuss other weaknesses 
of Wright's metaphor. I have argued elsewhere (Gavrilets 1997a) that 
his metaphor of rugged adaptive landscapes with its
emphasis on adaptive peaks and valleys is, to a large degree, a reflection
of the three-dimensional world we live in. Both genotypes and phenotypes of
biological organisms differ in numerous characteristics, and,
thus, the dimension of ``real'' adaptive landscapes is much larger than three.
Properties of multidimensional adaptive landscapes are very different from 
those of low dimension. Consequently, it may be misleading to use
three-dimensional analogies implicit in the metaphor of rugged adaptive 
landscapes in a multidimensional context. I believe that understanding
speciation requires a different metaphor. 

\begin{figure}[h]

\begin{center}
\scalebox{0.35}{\includegraphics[1.5in,1.in][8in,10in]{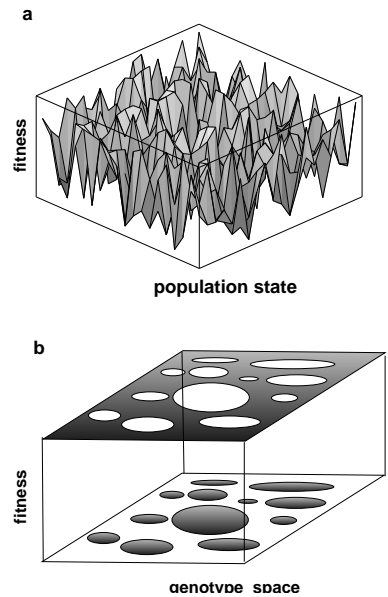}}
\end{center}

\caption{ {\bf Adaptive landscapes.
a. A rugged adaptive landscape. The main 
emphasis is on ``peaks'' and ``valleys'' corresponding to different 
well-adapted and maladaptive genotypes. 
b. A holey adaptive landscape. 
The main emphasis is on clusters of well-fit genotypes 
that extend throughout the genotype space. All other genotypes are 
treated as ``holes''.}}
\end{figure}

{\em The metaphor of ``holey'' adaptive landscapes}.
An individual organism can be considered as a combination of genes. 
All possible 
combinations of genes form a genotype space (which, mathematically, can be
represented by a hypercube). In discussing the evolution of 
populations, it is useful to visualize each individual as a point in this 
genotype space. Accordingly, a population will be a cloud of points, and 
different populations (or species) will be represented by different clouds. 
Selection, mutation, recombination, random drift and other factors change 
the size, location and structure of these clouds. To construct an adaptive
landscape one assigns ``fitness'' to each genotype (or each pair of genotypes)
in genotype space. Different forms of selection 
and reproductive isolation can be treated within this conceptual framework.
For example, fitness can be a genotype's viability (in the case of
viability selection); it can be fertility or 
the probability of successful mating between a pair of genotypes (in the case
of fertility selection or premating isolation, respectively).
A finite population subject to mutation is likely to be represented by 
genotypes with fitnesses within a fitness band determined
by the balance of mutation, selection and random drift.  
A general property of adaptive landscapes with a very large number of
dimensions is that genotypes with fitnesses within a specified band form connected ``clusters'' that extend throughout the genotype space 
(Gavrilets 1997;
Gavrilets and Gravner 1997). Thus, populations can evolve and diverge along
bands of highly-fit genotypes without going across the states with a 
large number of low-fit genotypes (that is without crossing any adaptive
valleys).

The metaphor of ``holey'' adaptive landscapes puts special emphasis on these clusters of highly-fit
genotypes, disregarding fitness differences between them and treating all 
other genotypes as ``holes'' (Gavrilets 1997; Gavrilets
and Gravner 1997). The justification
for the latter is a belief that selection will be effective in moving 
the population away from these areas of genotype space on a time scale 
that is much faster than the time scale for speciation. Accordingly, 
microevolution and local adaptation can be viewed as the
climbing of the population from a ``hole'' towards the holey adaptive
landscape, whereas macroevolution can be viewed as a movement
of the population along the holey landscape with speciation taking place
when the diverging populations come to be on opposite sides of a ``hole'' 
in the adaptive landscape.  In this scenario, there
is no need to cross any ``adaptive valleys''; reproductive 
isolation between populations evolves as an inevitable side effect 
of accumulating different mutations. The metaphor of holey adaptive
landscapes can be traced to a verbal two-locus two-allele model of
reproductive isolation proposed by Dobzhansky (1937) and similar
ideas discussed by Bateson (cited in Orr, 1997), Muller (1942),  
Maynard Smith (1970, 1983), Nei (1976), Barton and Charlesworth
(1984), Kondrashov and Mina (1986). For more discussion of this metaphor
see Gavrilets (1997ab) and Gavrilets and Gravner (1997). Orr (1995) and
Gavrilets (1997a)
gives a summary of relevant experimental evidence.  
The metaphor of ``holey'' adaptive landscapes is illustrated graphically
in Figure 1b.

{\em Mathematical models for holey adaptive landscapes.}
Here I briefly review previously published work on the evolutionary
dynamics on ``holey'' adaptive landscapes. 
Nei (1976) and Wills (1977) were the first to present formal analyses of
the Dobzhansky model.
Nei et al. (1983) studied one- and two-locus multi-allele models with step-wise
mutations and considered both postmating and premating reproductive isolation. 
In their models genotypes were reproductively isolated if they were 
different by more than 1 or 2 mutational steps. In these situations, 
speciation was very slow.
They conjectured, however, that increasing the number of loci
may significantly increase the rate of speciation. 
Bengtsson and Christiansen (1983) presented a deterministic analysis of 
mutation-selection balance in the Dobzhansky model.
Bengtsson (1985), Barton and Bengtsson (1986) and Gavrilets (1997b) analyzed the properties of hybrid zones arising under Dobzhansky-type
epistatic selection. Wagner et al. (1994)
considered a two-locus, two-allele model of stabilizing selection acting
on an epistatic character. For a specific set of parameters, the interaction 
of epistasis in the trait and stabilizing selection on the trait resulted
in a fitness ``ridge''. The existence of this ridge simplified stochastic
transitions between alternative equilibria. Gavrilets and Hastings (1996)
formulated a series of two- and three-locus Dobzhansky-type viability selection
models as
well as models for selection on polygenic characters. They studied
these models in the context of founder effect speciation and
noticed that the existence of ridges in the adaptive landscape
made stochastic divergence much more 
plausible. Similar conclusions were reached by Gavrilets and Boake (1998)
who studied the effects of premating reproductive isolation on the 
plausibility of founder effect speciation.
Higgs and Derrida (1991, 1992) proposed a model where the
probability of mating between two haploid individuals is a decreasing 
function of the proportion of loci at which they are different. 
Here, any two sufficiently different genotypes can be considered as 
sitting on opposite sides of a hole in a holey adaptive landscape.
These authors as well as Manzo and Peliti
(1994) studied this model numerically assuming that the number of the
loci is infinite, the loci are unlinked and highly mutable, and 
mating is preferential.
Orr (1995) and Orr and Orr (1996) studied speciation in
a series of models in which viability of a diploid organism depends on 
the number of heterozygous loci. All these papers {\em postulated} the 
existence of ridges of highly-fit genotypes.
Gavrilets and Gravner (1997) studied a general class of multilocus selection
models and showed the existence of ridges to be inevitable under fairly 
general conditions.
Independently, a similar conclusion was reached in Reidys et al. (1997).
Most previous studies of the dynamics of speciation on holey adaptive
landscapes were numerical. To develop a dynamical theory of speciation
it is desirable to have a simple model that can be treated analytically.

\section{The Model}
I consider finite populations of haploid individuals with discrete,
non-overlapping generations. I assume that reproduction involves gene
exchange (amphimixis) between individuals. The restriction to haploids is
for algebraic simplicity. Models for diploids will be discussed later.
Individuals are different 
with respect to $L$ possibly linked diallelic loci. Without any loss 
of generality each individual's genotype can be represented as a 
sequence of 0's 
and 1's. Let $l^{\alpha}=(l_1^{\alpha},...,l_L^{\alpha})$ where 
$l_i^{\alpha}$ is equal to 0 or 1, be such a sequence for an individual
$\alpha$. 
In standard population genetics models, the population state is usually 
described in terms of gamete frequencies. In systems with many loci such an
approach is not practical.
For instance, with 10 diallelic loci there are $2^{10}$ different gametes. 
Thus, one would need to analyze more than 1000 coupled equations. Another
complication follows from the fact that even in very large populations with
hundreds of thousands individuals each specific genotype is represented
only by a small number of copies or is not represented at all. Thus, the 
notion of a gamete frequency in multilocus evolution might be very difficult 
to justify. Here, I will be interested in 
the levels of genetic variation within subpopulations and genetic divergence
between subpopulations. Both can be characterized in terms of genetic distance
$d$ defined as the number of loci at which two individuals are different. 
More formally, the genetic distance $d^{\alpha \beta}$ between individuals
$\alpha$ and $\beta$ is
   \begin{equation} \label{distance}
          d^{\alpha \beta} = \sum_{i=1}^L (l_i^{\alpha}-l_i^{\beta})^2.
   \end{equation} 
Genetic distance $d$ is the standard Hamming distance. It is analogous to the
number of segregating sites in a sample of two gametes, which is widely used 
in molecular evolutionary genetics (Li, 1997), and to the number of 
heterozygous loci in a diploid organism. Genetic distance $d$
is also closely related to the notion of the overlap $q$ between
two sequences, $d=\frac{L}{2}(1-q)$, which is commonly used in 
statistical physics (e.g., Derrida and Peliti 1991). I model
the expected dynamics of average genetic distances within and between 
populations, using $D_w$ for the former and $D_b$ for the latter.

I assume that reproductive isolation is caused by cumulative genetic 
change. I will use a very simple symmetric model which is closely 
related to the models discussed above and which allows one to treat both 
pre- and postmating isolation within the same framework.
I posit that an encounter of two individuals can result in viable and 
fecund offspring only if the individuals are different 
at no more than $K$ loci. Otherwise the individuals do not mate 
(premating reproductive isolation) or these offspring are inviable or 
sterile (postmating reproductive isolation). More formally, I assign
``fitness'' $w$ to each pair of individuals depending on the genetic
distance $d$ between them
    \begin{equation} \label{thres}
         w(d) = \left\{ \begin{array}{cc}
                        1 & \mbox{for $d\leq K$},\\
                        0 & \mbox{for $d>K$}.
                       \end{array}
                \right.
    \end{equation}
(see Appendix for an outline of more complicated approaches).
In this formulation, any two genotypes different at more
than $K$ loci can be conceptualized as sitting on opposite sides of a hole
in a holey adaptive landscape (cf. Higgs and Derrida 1991, 1992). 
At the same time, a population can evolve
to any reproductively isolated state by a chain of single locus
substitutions. The adaptive landscape corresponding to this model is both
``holey'' and ``correlated''. The latter means that the probability that
two genotypes are reproductively isolated correlates with the genetic
distance between them. In Nei et al. (1983) and Gavrilets and Boake (1998)
models individuals separated 
by more than one mutational step were reproductively isolated which 
corresponds to $K=1$. The neutral case (no reproductive isolation) 
corresponds to $K$ equal to the number of loci.

The mathematical model presented above was interpreted as describing
sexual haploid populations with fitnesses assigned to
pairs of individuals depending on the genetic distance between them. 
However, there is an alternative interpretation in that the model 
describes randomly mating diploid populations. In the diploid case,
the genetic distance (\ref{distance}) between the two gametes forming
an individual is
equivalent to individual's heterozygosity, and fitness function
(\ref{thres}) specifies fitness as a function of individual heterozygosity.
Therefore, most conclusions of this paper will also be applicable to 
situations when post-mating reproductive isolation is in the form of 
reduced (or zero) viability of hybrids due to incompatibility of the genes 
they receive from their parents (Wu an Palopoli 1994).

\section{Dynamics in the neutral case} 
Before developing a theory for the dynamics of speciation in 
the above model, it is illuminating 
to start with the neutral case. Here I summarize some relevant results 
that are presented in (or follow directly from) classical papers 
(Watterson 1975; Li 1976; Slatkin 1987; Strobeck 1987). 
Let $\mu$ be the probability of mutation per locus per generation. 
The approximations below assume that mating is random,
the number of loci $L$ is large, 
but $\mu$ is very small so that the probability of mutation 
per individual per generation $v \equiv L\mu<<1$. The migration rate $m$
and the inverse of the population size $1/N$ are small as well.

{\em Genetic variation within an isolated population}.
Let us consider an isolated population of size
$N_T$. The expected change in the average genetic distance within the 
population per generation is
        \begin{equation} \label{Dwith}
             \Delta D_w = 2v - \frac{D_w}{N_T},
        \end{equation}
where the first term in the right-hand side is the contribution of mutation
whereas the second term is the random drift reduction of $D_w$. Asymptotically,
a mutation-drift equilibrium is reached with
        \begin{equation} \label{theta}
             D_w^* = \theta \equiv 2vN_T.
        \end{equation}

{\em Genetic divergence between isolated populations}.
Let us consider several isolated populations of arbitrary size. The probability
that a specific mutation gets fixed in a population is one over the
population size. Different mutations will get 
fixed in different populations resulting in their genetic divergence. 
The average genetic distance between any two of them increases with the 
rate equal twice the mutation rate per gamete
        \begin{equation} \label{speed}
             \Delta D_b = 2v.
        \end{equation}
The rate of neutral divergence does not depend on population sizes. In
particular, it is the same independent of whether there are many small
populations or a few large populations. Because the number of loci $L$ is 
finite, an indefinite increase of $D_w$, which
is implied by equation (\ref{speed}) is impossible. This equation as well 
as equation (\ref{Dwith}) above and equations (\ref{neut_dyn}) below 
approximate the dynamics when genetic distances $D_w$ and $D_b$ are small
relative to the number of loci $L$. To treat the general case, one has to
substitute $v$ for $v(1-2D_w/L)$ in equations (\ref{Dwith}) and
(\ref{neut_dyn-a}) and for $v(1-2D_b/L)$ in equations (\ref{speed}) and
(\ref{neut_dyn-b}). With a finite number of loci, the genetic distance $D_b$
between isolated populations approaches $L/2$ asymptotically.

{\em Subdivided populations}.
The effect of migration on the average genetic distances depends on the
spatial structure of populations.
Assume that a population of size $N_T$ is subdivided into $n$  
subpopulations of size $N=N_T/n$ and that a proportion $m>0$ of individuals migrate to any of the other $n-1$ subpopulations. The
expected changes in the average genetic distances within and between 
subpopulations are
     \begin{subequations} \label{neut_dyn}
        \begin{eqalignno}
            \Delta D_w & =  2v + 2m(D_b-D_w) - \frac{D_w}{N}, \label{neut_dyn-a}\\
            \Delta D_b & =  2v + \frac{2m}{n-1}\ (D_w-D_b). \label{neut_dyn-b}
        \end{eqalignno}
     \end{subequations}
Equations (\ref{neut_dyn}) assume that $v, m$ and $1/N$ are small.
Asymptotically, a mutation-migration-drift equilibrium is reached with
      \begin{subequations}
        \begin{eqalignno}
            D_w^* & = \theta,\\
            D_b^* & = \theta +  (n-1)\ \frac{2v}{m},
        \end{eqalignno}
     \end{subequations}
where $\theta$ is given by equation (\ref{theta}).
The average genetic distance within a subpopulation (of size $N$) does
not depend on the number of subpopulations $n$ and migration rate $m$ and 
is the same as is expected in a single population with size $N_T$.
The average genetic distance between sub-populations increases with the
population subdivision and decreasing migration. 
Figure 2 illustrates the dynamics of neutral divergence in a system of
two subpopulations.

\begin{figure}[tbh]

\begin{center}
\scalebox{0.35}{\includegraphics[1.5in,3in][8in,7.5in]{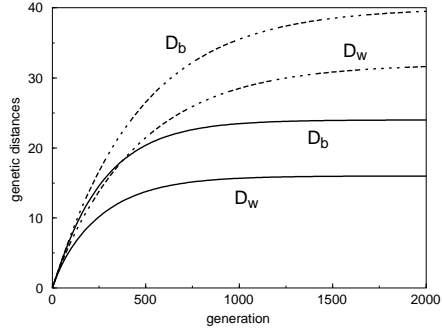}}
\end{center}

\caption{ {\small Dynamics of $D_w$ and $D_b$ in the neutral case.
Population size $N=100$ (solid lines) and $N=200$ (dashed lines).  
The rate of migration is $m=0.01$, the mutation rate per individual
is $v=.0384$. Initially, $D_w=D_b=0$.}}
\end{figure}

{\em Peripheral population}.
Assume a ``peripheral'' population of size $N$ is
receiving migrants from a very large ``main'' population. Genetic 
variation in the main population is assumed to be constant (and not 
influenced by migration from the peripheral population). The
expected changes in
the average genetic distances within the peripheral population, $D_w$, and
between the peripheral and main populations, $D_b$ are   
     \begin{subequations} \label{Li}
        \begin{eqalignno}
            \Delta D_w & =  2v + 2m(D_b-D_w) - \frac{D_w}{N},\\
            \Delta D_b & =  v + m(D_0-D_b) 
        \end{eqalignno}
     \end{subequations}
where $D_0$ is the average genetic distance within the main population
and $m$ is the proportion of individuals in the peripheral population
replaced by migrants from the main population. Asymptotically, a
mutation-migration-drift equilibrium is reached with
      \begin{subequations}
        \begin{eqalignno}
            D_w^* & = \frac{2Nm}{1+2Nm}\ D_0,\\
            D_b^* & = D_0 + \frac{v}{m},
        \end{eqalignno}
     \end{subequations}
where the former equation assumes that genetic variation in the main
population is sufficiently large ($D_0>>vN$) and the number of migrants, $Nm$,
is not too small.
The average genetic distance within the peripheral population is always larger
than that for an isolated population of its size ($D_w^*>2vN$).
If the number of migrants is large ($Nm>>1$), the average genetic distance
within the peripheral population is about the same as in the ``main''
population. 

\section{Dynamics with reproductive isolation}

The main feature of the model for reproductive isolation introduced above 
and other models of
holey adaptive landscapes is the existence of chains of equally-fit 
combinations of genes separated by single substitutions
that extend throughout the genotype space.
These chains can be though of as "neutral paths" in the 
adaptive landscape.
It is important to realize, however, that the existence of ``holes'' in
a holey adaptive landscape makes the actual dynamics of genetic
divergence not neutral. 
In this section I summarize some analytical results on the evolutionary dynamics
in the case of reproductive isolation described by equation (\ref{thres}).
Gavrilets et al. (1998) have studied the possibilities for speciation in this
model numerically. 
Details of the analytical methods used are outlined in the Appendix.
To derive the dynamic equations below, I have used the same assumptions
as described above at the beginning of the section on the neural case
substituting the assumption of random mating for the assumption of random
encounters. In addition, I have assumed that
the distributions of genetic distances both within and between populations
are Poisson. There are several sets of approximations resulting in Poisson
distribution of genetic distances. In the present contest, the weakest set
seems to be the assumption that genetic variation at each locus is small
most of the time (rare-alleles approximation) and that the population is approximately at 
linkage equilibrium. 
These assumptions are standard in analyzing the dynamics of multilocus systems under the joint action of selection, mutation, and random drift 
(e.g., Barton 1986; Barton and Turelli 1987; B\"{u}rger et al. 1989;
Gavrilets and de Jong 1993). The fit of individual-based 
simulations with analytic predictions is satisfactory both at qualitative 
and quantitative levels (see below and Gavrilets et al. 1998).

{\em Genetic variation within an isolated population}.
After the population becomes polymorphic at $K$ loci, new mutations
are selected against when rare because individuals carrying them have a 
reduced probability of producing viable and fecund offspring.
Selection experienced by individual loci underlying reproductive isolation
is frequency-dependent (and is similar to that arising in the case of
underdominant selection on a diploid locus). 
The change in $D_w$ per generation in an isolated population of size $N$ is
approximately 
     \begin{equation} \label{single}
          \Delta D_w = -s D_w + 2v - \frac{D_w}{N},  
     \end{equation}
where
     \begin{equation} \label{S}
          s =    \frac{e^{-D_w} D_w^K}{\Gamma(K+1,D_w)}
     \end{equation}
and $\Gamma(x,y)$ is an incomplete gamma function (e.g., Gradshteyn and
Ryzhik, 1994).
The value of $D^*_w$ at the mutation-drift-selection equilibrium can be found
by equating the right-hand side of (\ref{single}) with zero and solving 
for $D_w$.
Figure 3a illustrates the dependence of $D_w^*$ on the parameters of the model.
This figure indicates that the equilibrium values of $D_w$ are close to the
corresponding neutral predictions (\ref{theta}) if $K$ is larger than 2 
or 3
times $\theta$  (where $\theta=2Nv$ is the average genetic distance within
a finite population in the neutral case).
Figure 3b gives the values of the effective selection coefficient $s$.
With moderately large $K$ (that is with $K \geq 10$), $s$ is very small.
The effective selection coefficient $s$ can also be thought of as the strength
of induced selection on each locus underlying reproductive isolation. 
Figure 3b shows that very strong selection on the whole genotype (implied 
by the existence of complete reproductive isolation at finite values of
$K$) results in very weak selection at the level of individual loci.

\begin{figure}[tbh]

\begin{center}
\scalebox{0.35}{\includegraphics[1.5in,1.5in][8in,9.5in]{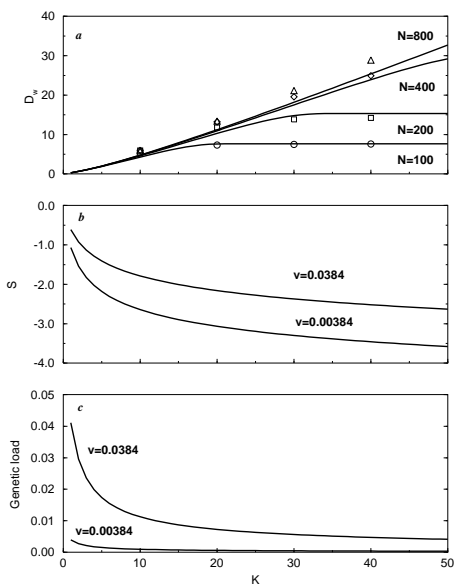}}
\end{center}
\caption{ {\small (a) Average genetic distance $D_w$ maintained by 
mutation-selection-drift balance in an isolated 
population of size $N$ as a function of $K$ for $v=0.0384$. 
The circles, squares, diamonds and triangles give estimates from 
individual-based simulations for $N=100, 200, 400$ and $800$,
respectively (thirty runs for each parameter configuration).
(b) Effective selection coefficient $s$ in the case of infinite population
size for two values of $v$. (c) The genetic load $1-\overline{w}$ for
parameters values as in Figure b.}}
\end{figure}

The mean fitness of the population, $\overline{w}_w$, can be defined as the 
proportion of pairs of individuals that can mate and produce fertile and 
viable offspring (cf., Nei et al. 1983). For a population with an 
average genetic distance $D_w$,
    \begin{equation} \label{wmean}
         \overline{w}_w = \frac{\Gamma(K+1,D_w)}{\Gamma(K+1)},
    \end{equation}
where $\Gamma(x+1)$ is a gamma function (e.g., Gradshteyn and
Ryzhik, 1994. For integer $x$, $\Gamma(x+1)=x!$).
Figure 3c shows that in spite of relatively high levels of genetic variation
maintained in the population, the genetic load (that is the proportion
of reproductively isolated pairs of individuals, $1-\overline{w}_w$) is very 
low. This seems to be a general property of holey adaptive landscapes 
(cf. Wills 1977; Bengtsson and Christiansen 1983).

{\em Genetic divergence between isolated populations}.
Even after the genetic distance within an isolated population has reached
an equilibrium level, the population keeps evolving as different mutations
get fixed. As a consequence, isolated populations will continuously diverge
genetically. The asymptotic rate of divergence of two isolated populations 
of size $N$ each is
  \begin{subequations} \label{isol}
    \begin{equation} \label{isol-a}
       \Delta D_b = 2 v R, 
    \end{equation}
where 
    \begin{equation}  \label{isol-b}
      R = \frac{2 e^{-S}\sqrt{S}}{\sqrt{\pi}\ erf( \sqrt{S})} 
    \end{equation}
  \end{subequations}
is the rate of divergence relative to the neutral case.
Here, $S=Ns/2$, $s$ is defined by equation (\ref{S}) with 
$D_w$ corresponding to the 
mutation-selection-drift equilibrium, and $erf(x)$ is the error function
($ = 2/\sqrt{\pi}\ \int_0^x exp(-y^2)dy$). In the neutral case, $s=0$,
$U=1/N$ and equation (\ref{isol-a}) reduces to equation (\ref{speed}).
Figure 4 illustrates the dependence of the relative rate of divergence $R$
on model parameters. In the neutral case, the rate of genetic divergence 
$\Delta D_b$ does not depend on the population size (equation \ref{speed}). 
In contrast, with reproductive isolation the rate
of divergence decreases with increasing population size. After the population becomes polymorphic at $K$ loci, new mutations
are selected against when rare. Genetic drift operating in finite populations
overcomes the effect of selection and allows genetic divergence. 
For example with $K=20$ and $v=.0384$, a population of size $N=800$ will 
accumulate about 5 substitutions per
1000 generations. A few thousand generations will be sufficient for $D_b$
to exceed $K$ significantly. In contrast, very large randomly mating 
populations will diverge very slowly.

\begin{figure}[tbh]

\begin{center}
\scalebox{0.35}{\includegraphics[1.5in,1.5in][8in,9.in]{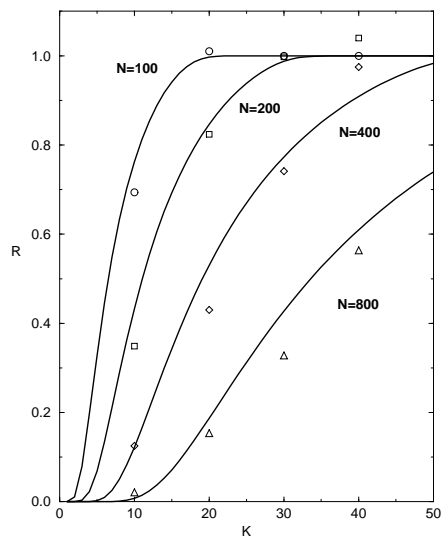}}
\end{center}
\caption{ {\small The rate of difergence $R$ relative to the neutral case
in an isolated population of size $N$ as a function of $K$ for $v=.0384$.
The circles, squares, diamonds and triangles give estimates from 
individual-based simulations for $N=100, 200, 400$ and $800$,
respectively (thirty runs for each parameter configuration).}}
\end{figure}

Figure 4 indicates that the rate of substitutions is close to the
corresponding neutral predictions if $K$ is larger than 2-3 times
$\theta$ ($\theta=2Nv$).
Note that as in the neutral case considered above, an implicit assumption in equation (\ref{isol}) is 
that genetic distance $D_b$ is small relative to the number of loci $L$. 
In the general case, $\Delta D_b=2v(1-2D_b/L)R$ and $D_b$ approaches $L/2$ asymptotically.

At what moment can the two diverging population be considered as two
different species? The answer obviously depends on what one means by a species.
Let us say that the two populations are different species if the proportion,
$\overline{w}_b$,
of encounters between individuals from different populations that can
result in mating and viable and fertile offspring is less than a small number
$\gamma$. (This definition uses the biological species concept.) During
initial stages of divergence, this proportion can be approximated by the 
right-hand side of equation (\ref{wmean}) with
$D_b$ taking the place of $D_w$. Figure 5 shows the minimum genetic distance
between populations required for speciation as a function of $K$ for
several values of $\gamma$. One can see that a genetic distance between the
populations on the order of 2 or 3 times $K$  will be sufficient for the
status of separate ``biological'' species. Note that there is very little 
effect of the magnitude of $\gamma$.

\begin{figure}[tbh]

\begin{center}
\scalebox{0.35}{\includegraphics[1.5in,1.5in][8in,9.in]{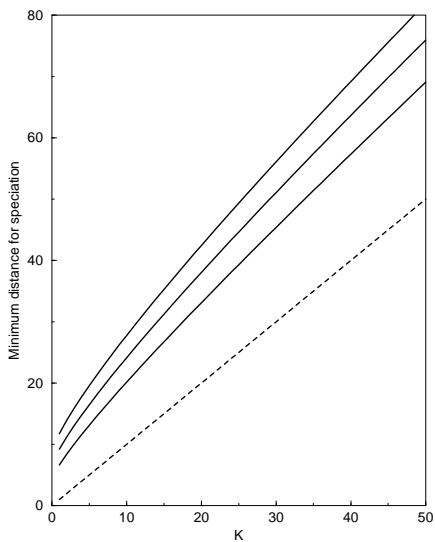}}
\end{center}
\caption{ {\small Minimum genetic distance between populations for speciation
for $\gamma=.0001, .001$ and
$.01$ (solid lines from top to bottom). Also shown is the diagonal $D=K$
(dashed line)}}
\end{figure}

{\em Speciation in a subdivided population}. 
In the deterministic limit (that is with $N_T \rightarrow \infty$), the genetic 
variation of a subdivided population can be maintained  by migration. 
This can happen if initially alternative alleles are close to fixation
in different subpopulations and selection is 
sufficiently strong relative to migration (e.g., Karlin and McGregor 1972). 
Let $k$ be the number of loci at which alternative alleles are close to
fixation in different subpopulations. 
Respectively, $L-k$ will be the number of loci at which the same allele is 
close to fixation in different subpopulations. In the deterministic limit,
$k$ does not change. In the $n$-island model the dynamics of $D_w$ and 
$D_b$ are described by equations
  \begin{subequations} \label{migr}
     \begin{eqalignno}
        \Delta D_w & = - s D_w + 2v + 2m_e (D_b-D_w),\\ \label{migr-a}
        \Delta D_b & = - s (D_b-k) + 2v + \frac{2m_e}{n-1} (D_w-D_b), \label{migr-b}
     \end{eqalignno}
  \end{subequations}
where $s$ is defined by equation (\ref{S}), and the ``effective'' migration 
rate
   \begin{equation} \label{me}
     m_e = m\ \frac{ \overline{w}_b}{\overline{w}_w}
   \end{equation}
if $k\leq K$ and $m_e=0$ otherwise. Here, $\overline{w}_w$ is given by 
equation (\ref{wmean}) above whereas 
$\overline{w}_b=\Gamma(K+1-k,D_b-k)/\Gamma(K+1-k)$ is
the probability that two randomly chosen individuals from different populations
are not reproductively isolated.
The effective migration rate $m_e$ can be thought of as half the probability of
mating between individuals from different subpopulations. With no reproductive
isolation (with very large $K$) or no genetic divergence between subpopulations
(with $D_b \approx D_w, k=0$), the effective migration rate is equal to 
the actual migration rate ($m_e=m$).
Comparing equations (\ref{migr}) with their neutral analogs (\ref{Li}) 
shows that reproductive
isolation results in two effects. First, it directly reduces genetic
variation within subpopulations and genetic divergence between subpopulations.
These effects are described by the first terms in the right-hand side of
equations (\ref{migr}). 
Also, reproductive isolation reduces the gene flow between populations.
Given that $D_b \geq D_w$, then $m_e\leq m$ reflecting the fact that 
genes brought
by migrants have a smaller probability of being incorporated in the resident
population. In the deterministic limit, both $D_w$ and $D_b$ always
evolve to finite equilibrium values.

Random genetic drift results in two effects. First, it reduces the genetic 
variation within sub-populations by the amount $D_w/N$. The dynamic
equation for $D_w$ becomes
      \begin{equation} \label{corrected}
         \Delta D_w = - s D_w + 2v + 2m_e (D_b-D_w)-\frac{ D_w}{N}.
      \end{equation}
Second, genetic drift might change $k$. The expected change in $k$ can be
approximated as
    \begin{equation} \label{k}
        \Delta k = 2v R 2^{-Nm_e} - 2km_e R (e/2)^{Nm_e},
    \end{equation}
where $R$ is given by equation (\ref{isol-b}). The first term in the right-hand
side of equation (\ref{k}) can be thought of as the rate at which an allele
that is initially rare in both sub-populations becomes close to fixation in 
one of the subpopulations. This rate was found by Lande (1979) using a
diffusion approximation and assuming that migration is weak. The second term 
is the rate at which the loci 
with different alleles initially close to fixation in different subpopulations
become fixed for the same allele in both of them. To find this term I used
Barton and Rouhani's (1987) method.

Depending on parameter values and initial conditions there are two different
dynamic regimes. In the first regime, both $D_w$ and $D_b$
evolve to finite values, which are smaller than those
in the neutral case (and which are much smaller than the number of loci $L$).
Here, selection (reproductive isolation) reduces 
genetic divergence both within and between subpopulations. 
In the second regime, $D_w$ stays small (relative to $L$) 
whereas $D_b$ increases effectively indefinitely (to values
order $L/2$). Here, selection (reproductive isolation) reduces the effective migration rate to zero resulting in speciation.
These dynamics can be understood in the following way.
Changes in $D_w$ and $D_b$ induced by selection 
are expected to happen on a faster time scale than changes in $k$ induced
by random genetic drift. Thus,
$D_w$ and $D_b$ values should be close to the equilibrium values predicted by
equations (\ref{migr-b},\ref{corrected}) when $k$ is treated as a constant.
The dynamic behavior depends on whether $k$ reaches a finite equilibrium
value or keeps increasing. In the latter case, the effective migration rate
$m_e$ reduces to zero, the rate of change of $k$ approaches $2vR$ with $D_b$
increasing at the same rate (cf. equation \ref{isol-b}).

\begin{figure}[tbh]

\begin{center}
\scalebox{0.4}{\includegraphics[.75in,1.in][8in,9.5in]{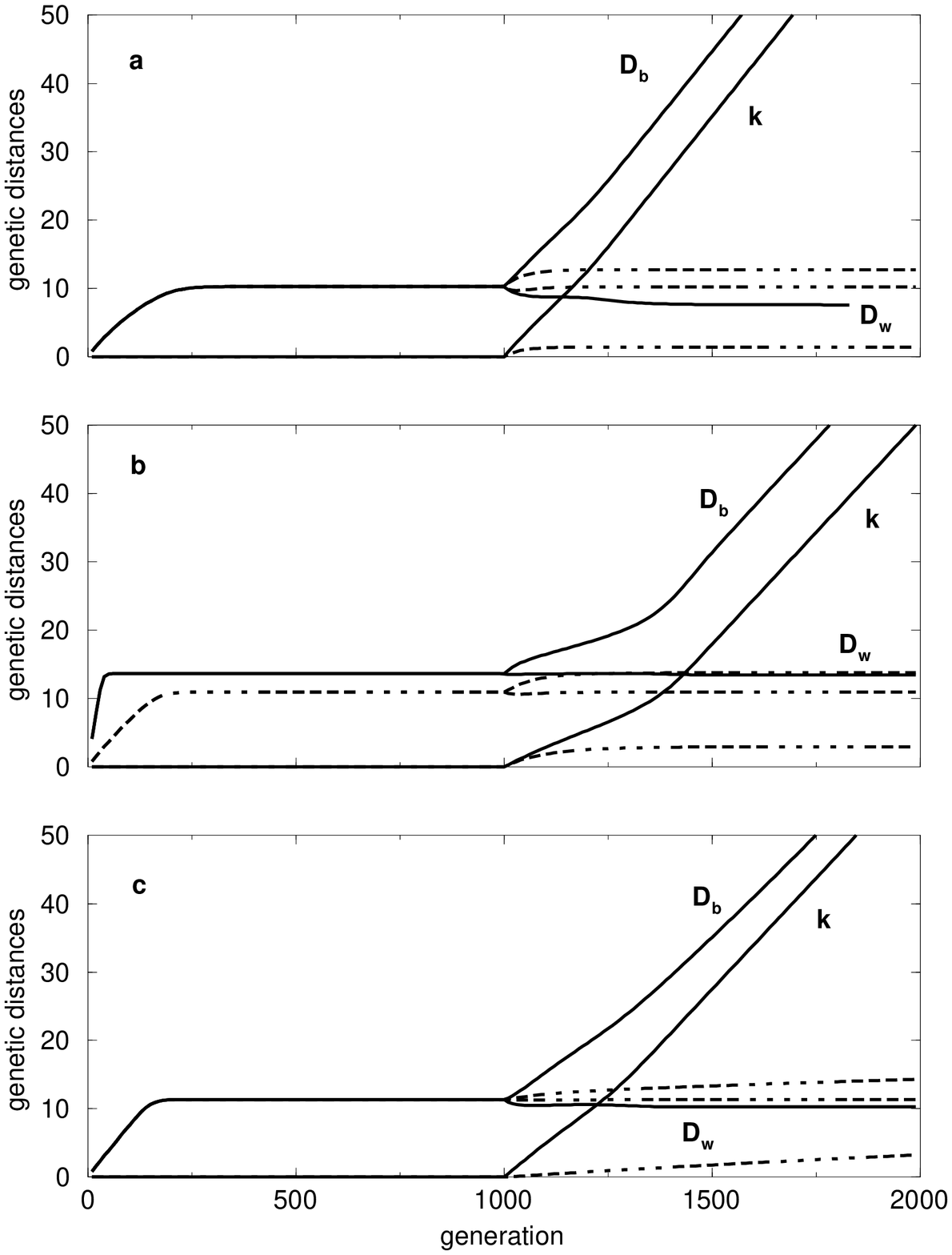}}
\end{center}
\caption{ {\small Dynamics of speciation in a subdivided population.
Unless specified otherwise, $K=20, v=0.0384, n=2$. a. Effects of migration 
rate. Stronger migration, $m=0.01$ (dashed lines; no speciation), and 
weaker migration, $m=0.001$ (solid lines; speciation). Total population size 
$N_T=200$. b. Effects of mutation rate. Weaker mutation, $v=0.0384$ (dashed 
lines; no speciation), and stronger mutation, $v=5\times 0.0768$ (solid lines; 
speciation). Other parameters: $N_T=400$, $m=0.005$. c. Effect of population 
subdivision. $n=2$ subpopulations (dashed lines; no speciation) and $n=4$ 
subpopulations (solid lines; speciation). Other parameters: $N_T=800$, 
$m=0.0033$. Dashed lines represent $D_b$ (top line), $D_w$ (middle line) 
and $k$ (bottom line), respectively. During the first 1000 generations 
there are no restrictions on migration.
}}
\end{figure}

Figure 6 illustrates the dynamics observed by numerically iterating the model equations. 
The iterations started with all $N$ individuals identical. During the first
1000 generations there were no restrictions on  migration  
between subpopulations and the whole population evolved as a single 
randomly mating unit (cf. Gavrilets et al. 1998). 
The average genetic distance within the population $D_w$ evolves according 
to equation (\ref{single}).
Starting with generation 1000,  restrictions on migration were introduced 
and the dynamics are described by equations (\ref{migr-b}-\ref{k}) afterwards.
After generation 1000, each of these figures has two sets of three curves corresponding to two
different values of the parameter(s) under consideration. The curves within
each set represent $D_b, D_w$ and $k$. With migration rate $m=0.01$, 
all these variables evolve towards finite equilibrium values (see Fig.6a)
whereas with a smaller migration rate ($m=0.001$), $D_b$ and $k$ increase
effectively indefinitely signifying that speciation has taken place. 
Thus, reducing
migration makes speciation more plausible. Figure 6b shows that increasing
mutation rate (from $v=0.0384$ to 5 times this value) has a similar effect. These two
figures describe the dynamics expected in a system of two subpopulations.
Figure 6c compares the dynamics observed in a population subdivided into
$2$ and $4$ subpopulations. This figure shows that increasing population
subdivision makes speciation more plausible. 
Note that the process of genetic divergence described in Fig.6c results in a
simultaneous emergence of 4 species.     
In the cases where speciation takes place (as signified by continuous 
increase in the genetic distance between subpopulations), 
the curves representing $D_b$ and $k$ are parallel meaning that
asymptotically the genetic divergence is due to fixation of different 
mutations in different subpopulations.
In the cases where speciation does not take place, $k$ is close to zero.

Altogether, at the qualitative level the results presented in Fig.6 
correspond to both biological intuition and the results of individual-based 
simulations in Gavrilets et al. (1998). 
At the quantitative level, there is a very good fit between simulations and
analytical predictions for levels of genetic variation maintained in 
subpopulations and the asymptotic rate of divergence between subpopulations.
However, the conditions for speciation as predicted by iterating equations 
(\ref{migr-b}-\ref{k}) appear to be more strict than those observed in
the individual-based simulations performed by Gavrilets et al. (1998). 
For example, for parameter values used in Figure 6c, no speciation in
a system of four subpopulations occurs if $m>0.0035$. In contrast, in
individual-based simulations speciation was observed for $m=0.01$ 
(Figure 3b in Gavrilets et al. 1998). The main reason for this disperancy 
seems to be an inadequacy of equation (\ref{k}) 
at moderate levels of migration (e.g. Lande 1979; Barton and Rouhani 1987).\\

\begin{figure}[tbh]

\begin{center}
\scalebox{0.4}{\includegraphics[.75in,1.in][8in,8.5in]{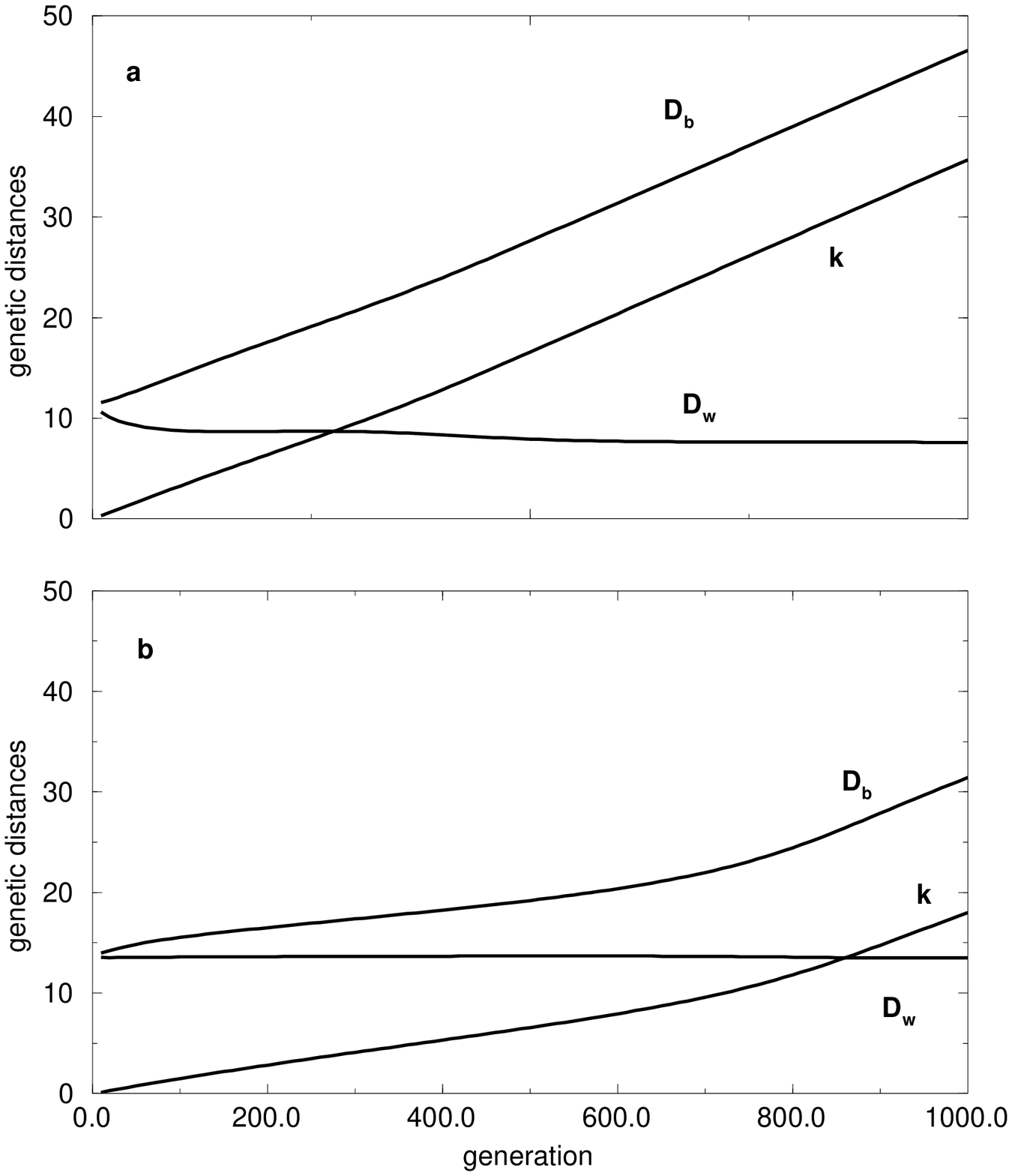}}
\end{center}
\caption{ {\small Dynamics of speciation in a peripheral population.
a. Speciation with $m=0.001; N=100; K=20; v=0.0384$ (cf. Fig.7a). 
b. Speciation with $m=0.005; N=200; K=20; v=5 \times 0.0768$ (cf. Fig.7b).
}}
\end{figure}

{\em Speciation in a peripheral population}.
Here I consider the case of a peripheral 
population of size $N$ receiving migrants from a very large main 
population. The dynamics of the average genetic
distance within the peripheral population, $D_w$, the average genetic
distance between the peripheral and main populations, $D_b$, and the number
of diverged loci $k$ are approximated by equations 

  \begin{subequations} \label{per}
     \begin{eqalignno}
        \Delta D_w  = &- s D_w + 2v + 2m_e (D_b-D_w) - \frac{D_w}{N},\\
        \Delta D_b  = &- \frac{s}{2}(D_b - k) + v + m_e(D_0-D_b), \\
        \Delta k    = & v R 2^{-Nm_e} - km_e R (e/2)^{Nm_e}.
     \end{eqalignno}
  \end{subequations}
Here $D_0$ is the average genetic distance within the main population.
Figure 7 illustrates the dynamics observed by numerically iterating equations
(\ref{per}). For $D_0$ I used mutation-selection balance values for a
very large isolated population predicted by equations (\ref{single},
\ref{S}). The initial values of $D_w$ and $D_b$ were equal to $D_0$.
The parameter values in Fig.7a and Fig.7b are the same as those
that resulted in speciation in Fig.6a and 6b, respectively. The outcome
of the dynamics is the same - speciation - but the rate of divergence is 
smaller than when all subpopulations are uniformly small.
This is apparent from the level of genetic distance between subpopulations
achieved after 1000 generations of divergence which are about twice as small in 
Figures 7a and 7b relative to those in Figures 6a and 6b.

\section{Discussion}

The theory developed above together with earlier numerical simulations
(see references above) show that rapid speciation is a plausible outcome 
of the evolutionary dynamics in subdivided populations. Here speciation 
is a consequence of two fundamental factors. The first factor is the existence 
of various and possibly significantly different highly-fit combinations of
genes underlying diverse solutions (genetical, ecological, behavioral, 
developmental etc.) to the problem of survival and reproduction. 
In multidimensional genotype space these combinations of genes tend to 
form connected clusters that extend throughout genotype space. 
At the same time these genotypes are not mutually compatible - they are
separated by ``holes''. The second factor is mutation pressure.
Because the population size is finite and the number of loci is very large
whereas the probability of a specific mutation is very small,
different mutations tend to appear (and increase in frequency) in different
subpopulations (cf., Barton 1989; Mani and Clarke 1990). 
Metaphorically speaking, mutation tends to tear apart the cloud
of points representing the population in genotype space. Combining 
genes from two different organisms in one offspring can counteract the
disruptive effect of mutation, keeping a single randomly mating population
together in genotype space. But restricting gene exchange as a consequence
of limited migration between subpopulations gives mutation a significant advantage. Eventually the population cloud will be broken
and smaller clouds representing the subpopulations will
drift apart in genotype space - an event representing speciation.
Given sufficient genetic divergence,
restoring migration to high levels will not return the system back
to the state of free gene exchange between subpopulations which now can be
considered as different species.
It is not necessary to invoke strong selection for local adaptation to
explain speciation in a subdivided population, as studied here, or
after a founder event (Gavrilets and Hastings 1996; Gavrilets and Boake 1998). Mutation is ubiquitous. Population size is never infinite and, 
thus, genetic drift is always present. Speciation as caused by mutation 
and random drift should represent a null model against which speciation 
as caused by local adaptation can be tested (cf., Nei 1976;
Lande 1976).

Unlike most previous models that concentrate only on some stages of speciation,
the model studied here describes the complete process of speciation from initiation until completion. I assumed that reproductive isolation is caused 
by cumulative genetic change. The model is described in terms of dynamic equations for the variables analogous to those used in molecular evolutionary
biology - the average genetic distances between and within subpopulations. 
Average genetic distances within (sub)populations always evolve towards finite 
equilibrium values. Depending on parameter values and initial conditions   
average genetic distances between subpopulations either converge to a finite 
equilibrium or increase effectively indefinitely. The former regime is interpreted as
no speciation. In the latter regime, three effects take place simultaneously:
(1) genetic distances between subpopulations significantly exceed genetic
distances within them, (2) encounters between individuals from different 
subpopulations do not result in viable and fertile offspring, (3) evolutionary
changes in a subpopulation do not affect other subpopulations.
Thus, subpopulations form separate genotypic clusters in genotype space,
become reproductively isolated and undertake changes as evolutionary 
independent units. This regime is interpreted as speciation according to 
any of the species concept common in the literature
(e.g., Mallet 1995; Claridge et al. 1997). 

The dynamic equations derived above describe the expected changes in the 
average genetic distances neglecting stochastic fluctuations around the 
expected values. The predicted dynamics have two clearly distinct regimes:
convergence towards a finite equilibrium (no speciation) or effectivly
indefinite
divergence (speciation). Stochastic fluctuations around the expected values,
which are present in natural populations (and individual-based simulations),
make the boundary between these two regimes less strict and may result in 
the population escaping the first regime and entering the second regime
after some time (see Gavrilets et al. 1998). 
My analysis has been based on approximations which are standard in studying
multilocus systems. I assumed that alleles are rare, that linkage disequilibrium, mutation and migration rates are small and used 
a theory developed by Lande (1979), Walsh (1982) and  Barton and Rouhani (1987)
for describing stochastic transitions driven by random genetic drift. 
The analytic theory presented 
here fits satisfactorily with the results of individual-based simulations.
The model can be used to evaluate qualitative effects of different factors
on the dynamics of speciation, the order of magnitude of parameters resulting 
in or preventing speciation, and the time scale involved.
According to both biological intuition and previous numerical simulations,
increasing mutation rate and decreasing migration promote speciation.
Increasing the number of loci has significantly increased the plausibility
of speciation relative to that in earlier models (Nei et al. 1983; Wagner
et al. 1995; Gavrilets and Hastings 1996). Note that the actual number of loci
influences the dynamics only through the mutation rate per gamete, $v$,
and parameter $K$. For realistic parameter values the time 
scale for speciation can be as short as a few thousands or even hundreds of
generations. This is compatible with rates observed in several cases
of rapid speciation in natural populations described recently (Schluter 
and McPhail 1992; Yampolsky et al. 1994; Johnson et al. 1996; 
McCune 1996, 1997) including the 
most spectacular case - the origin of hundreds of species of Lake Victoria 
cichlids in 12,000 years (Johnson et al. 1996).
The model has demonstrated the plausibility of speciation with
relatively low levels of both initial genetic variation and new genetic
variation introduced into the population each generation (both supplied 
by mutation). With higher levels of the former (as in laboratory
experiments on speciation, reviewed by Rice and Hostert 1993, and Templeton,
1996) or of the latter (for instance as a result of natural hybridization,
reviewed by Bullini 1994, Rieseberg 1995, Arnold 1997), the rate of 
speciation is expected to be even higher.

\subsection{Local adaptation and speciation}

The model analyzed above shows that rapid speciation in a subdivided population
can occur even without any differences between selection regimes operating
in different subpopulations (that is without selection for local adaptation). 
An important question is how genetic changes brought about by selection 
for local adaptation would affect the dynamics of speciation (e.g., 
del Solar 1966; Ayala et al. 1974; Kilias et al. 1980; Dodd 1989; Schluter 
1996; Givnish and Sytsma 1997). These effects will depend on whether the 
genes responsible for local adaptation are different from or are the same 
as the genes underlying reproductive isolation.

\begin{figure}[tbh]

\begin{center}
\scalebox{0.4}{\includegraphics[.75in,1.5in][8in,9.5in]{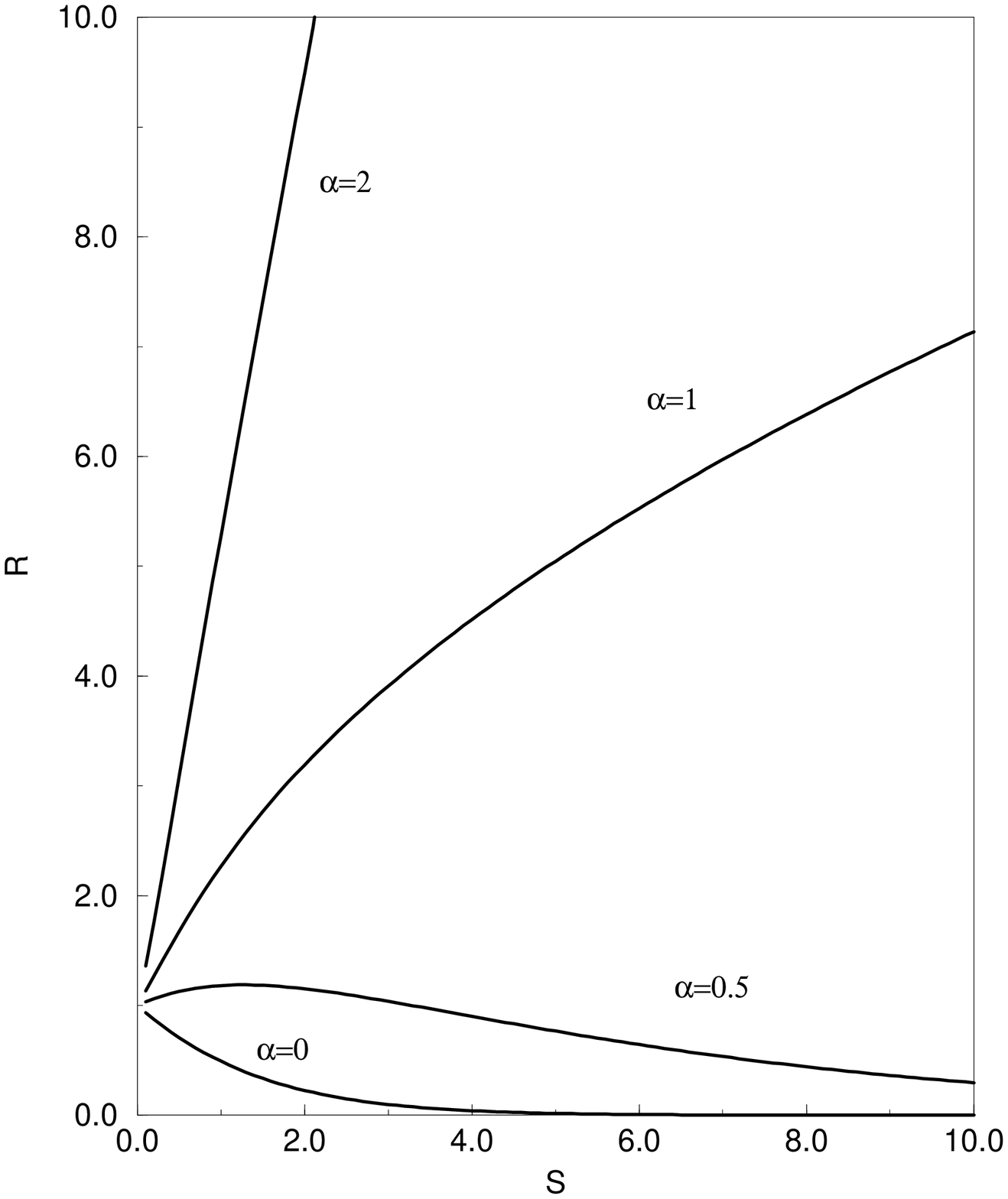}}
\end{center}
\caption{ {\small Relative rate of fixation in the case with local
adaptation.  
}}
\end{figure}

Assume first that the two sets of genes are completely different.
Let the strength of selection per locus induced by reproductive isolation be very small so that these loci can be considered as effectively neutral.
(For the model studied here, this seems to be the case if $K$ is larger
that 2-3 times $\theta$, where $\theta$ is the average genetic distance
maintained by mutation in a finite population in the neutral case.)
Then, Birky and Walsh's (1988) results tell us that the rate of
substitution in these loci will not be affected by selection on other loci
independently of linkage. However, given that reproductive isolation is a result
of genetic incompatibilities, the loci underlying reproductive
isolation will be under frequency-dependent selection against rare alleles
which is analogous to underdominant selection in diploid populations. 
Birky and Walsh (1988) have 
shown that linkage to advantageous alleles slightly increases the rate of 
fixation of detrimental mutations. This  suggests that selection on
linked loci will increase the rate of substitutions in the loci underlying
reproductive isolation and, thus, will promote speciation to some degree. 
No results seem
to be known on how linkage to advantageous alleles increases the rate of fixation of underdominant mutations or alleles experiencing frequency-dependent
selection. No quantitative predictions can be made here, but most likely
if the two sets of loci are not extremely tightly linked, effects of
selection for local adaptation on the rate of speciation will not be 
significant.

Assume now that the loci under consideration pleiotropically affect both
survival in a given environment and reproductive isolation. For instance, 
this may be the case if disruptive selection acts on habitat preferences 
which also define mating patterns (e.g., Rice 1984; Rice and Salt 1988) or 
if the probability of mating between individuals depends on the difference 
in their morphological traits which are under direct selection. 
Let $s_{LA}$ be the average strength of selection per locus induced by 
selection for local adaptation. Using Walsh's (1982) results, the relative 
rate of fixation of
new mutations in an isolated population of size $N$  can be approximated 
(see Appendix) as
       \begin{equation} \label{walsh}
             R = \frac{4 e^{-S(1-\alpha)^2} \sqrt{S}}
                      { \sqrt{\pi} \left[ erf(\sqrt{S}(1+\alpha)) + erf( \sqrt{S}(1-\alpha)) \right] },
       \end{equation}
where $S=Ns/2$, $s$ is the strength of selection per locus induced by
reproductive isolation, and $\alpha=s_{LA}/s$.
With $\alpha=0$, equation (\ref{walsh}) reduces to (\ref{isol-b}). 
Figure 8 illustrates the dependence of $R$ on $S$ and $\alpha$.
Increasing $\alpha$ always increases $R$. Thus, selection for local 
adaptation always increases the rate of substitutions and promotes speciation.
With sufficiently strong selection for local adaptation ($s_{LA}>s$), the
net effect of new alleles will be advantageous and their frequencies will
tend to increase even when rare. In the limit of large population size,
the probability of fixation is $2(s_{LA}-s)$. This is analogous to the 
classical results on the probability of survival of an advantageous mutant 
in a very large population (Haldane 1927; Walsh 1982). The rate of 
accumulation of genetic differences will be $2(s_{LA}-s)Nv$ and can be
significant. 
Very strong artificial selection for local adaptation has been 
shown to result in rapid evolution of reproductive isolation (e.g., 
del Solar 1966; Kilias et al. 1980; Dodd 1989). However, the changes
brought about by moderately strong artificial selection may not exceed 
those resulting from random genetic drift only (e.g. Ringo et al. 1985).

\subsection{Population subdivision and speciation}

In the models considered here, speciation is a by-product of fixation of
different alleles in different subpopulations.
It is well known that the rate of fixation of neutral alleles does not
depend on population size, that of advantageous alleles increases with
population size, and that of deleterious or underdominant alleles decreases
with population size (e.g., Gillespie 1991; Ohta 1992). At the level of individual loci, selection induced by reproductive isolation in the form
considered here is similar 
to underdominant selection (or frequency-dependent selection against 
rare alleles). Thus, in the absence of selection for local adaptation (or 
with independent loci controlling traits for local adaptation) decreasing population size will increase the rate of substitutions and promote speciation 
(see equation \ref{isol-b}). Effects of the population size on the 
plausibility of speciation will be similar even if the same loci control 
both reproductive isolation and locally beneficial traits given that selection 
for local adaptation is not too strong ($s_{LA}<s$, see equation \ref{walsh} 
and Fig.9). In all these cases, speciation will be driven by mutation and
random genetic drift and will be fastest if the population is subdivided 
into small subpopulations. This conclusion about the effect of 
population subdivision on the probability of speciation in Dobzhansky-type models differs from that of Orr and Orr (1996). They argued that the degree 
of population subdivision has no effect on the rate of speciation if 
speciation is caused by mutation and random drift. Orr and Orr did not 
consider the actual process of fixation of new mutations assuming that
it will be a simple neutral process. However, the existence
of holes in the adaptive landscape makes the process of substitution  
non-neutral and new mutations are selected against when rare. 
Such mutations are fixed more easily in smaller subpopulations.
For the discussion of the existing experimental evidence regarding effects
of random genetic drift on the plausibility of speciation see Rice and
Hostert (1993) and Templeton (1996).
The time scale for speciation is short meaning that restrictions on migration
between subpopulations do not need to be long lasting; several hundreds of 
generations may be sufficient for a significant divergence and evolution of 
reproductive isolation. It is quite possible that several new species will 
emerge from a highly subdivided population within a short period of time
(see above; Gavrilets et al. 1998). These theoretical conclusions are 
consistent with a verbal ``micro-allopatric'' model of speciation suggested 
for cichlid fishes in the East African Great Lakes (e.g., Reinthal and Meyer, 
1997). Hoelzer and Melnick (1994) have emphasized that the possibility of
simultaneous emergence of several new species should be incorporated more 
explicitly in the contemporary methods for reconstructing phylogenies.

If the same loci control both reproductive isolation and
locally beneficial traits and selection for local adaptation is sufficiently
strong ($s_{LA}>s$), increasing the population size will result 
in increasing the rate of substitutions (see Fig.9). In this case, speciation
will be driven by selection and will be 
fastest if the population is subdivided into a small number (say, two) of 
large subpopulations (Orr and Orr 1996) as implied by the vicariance
scenario (e.g. Wiley 1988).

Many species are thought to be represented by a few large populations and many 
smaller ``peripheral'' populations. Mayr (1963, 1982b) proposed the theory 
of peripatric speciation arguing that speciation
is typically initiated in small peripheral populations and he attributed a 
special role to genetic drift in this process. Gavrilets (1996) 
has shown that an invasion of a new adaptive combination of genes is most
successful if it is initiated in a peripheral population.
The results presented here
bear out Mayr's  argument(see Fig. 8). Small peripheral populations will
rapidly diverge genetically from the ``main'' large population and speciate.
Although differences in selection regimes between peripheral and main populations can accelerate divergence, random genetic drift will be the most
important factor. On the other hand, 
if a peripheral population is large enough and is under a selection regime
that is sufficiently different from the one operating in ``main'' 
populations, then disruption of gene flow can cause evolutionary divergence,
perhaps leading to rapid speciation, in the absence of contributions from
random genetic drift (Garcia-Ramos and Kirkpatrick 1997). 

Summarizing, large randomly mating populations will diverge genetically 
and speciate only if there is strong 
selection for local adaptation (for instance after a change in the 
environment). In contrast, small populations will diverge and speciate 
even without differences in selection regimes between them. 
Possibilities for speciation strongly depend on the geographic structure
of the population. Here, analysis was restricted to the island model and
the continent-island model.  Manzo and Peliti (1994) and Gavrilets et al. 
(1998) present numerical results for stepping-stone models.

\subsection{Relationship to other speciation models}

Using genetic distance (1) implies the equivalence of loci. 
A general case of non-equivalent loci can be described by introducing a $(L 
\times L)$ matrix $G=\{G_{ij}\}$ of weights and defining a generalized
 distance between individuals $\alpha$ and $\beta$ as
   \begin{equation} \label{gen_dist}
          d^{\alpha \beta} =(l^{\alpha}-l^{\beta})^T G (l^{\alpha}-l^{\beta}),
   \end{equation} 
where $l^{\alpha}$ and $l^{\beta}$ are vectors defining the corresponding
genotypes and superscript $T$ means transpose.
Considering haploid populations and premating isolation only, the model 
assumes that individuals can mate only if they are not too different 
genetically.  Here, the degree of reproductive isolation was controlled by 
cumulative genetic difference. 
However, using the generalized distance (\ref{gen_dist}) allows one to treat
models for reproductive isolation controlled by quantitative traits as well 
as models for sexual selection within the same framework (see Appendix).
The close relationship between the models of speciation as a consequence of
``quasi-neutral'' divergence along ridges in the adaptive landscapes and as
a consequence of sexual selection was already recognized
by Barton and Charlesworth (1984).

A fundamental reason for speciation on a holey adaptive landscape is mutation 
which tends to break the population into reproductively isolated pieces. Population 
subdivision and the resulting reduction in gene exchange facilitates this 
process.
Here, migration rates compatible with rapid speciation were small (that is 
speciation was allopatric or parapatric). An interesting question 
is whether speciation is possible with much higher migration rates.
In other words, is sympatric speciation by mutation and random genetic
drift on a holey adaptive landscape possible?
Numerical simulations of similar models of sympatric speciation
where mutation rates were higher (Higgs and Derrida 1991, 1992) or the 
time span studied was longer (Wu 1985) or the population size was smaller
(Gavrilets and Boake 1998) than here, provide an affirmative answer.
Adding disruptive selection due to either abiotic factors (say, different
resources) or biotic factors (competition) should create additional 
pressure on the population cloud which might result in rapid sympatric
speciation.

\subsection{Beyond holey landscapes}

Gavrilets and Gravner's (1997) results have suggested that clusters of 
well-fit genotypes that extend throughout genotype space are plausible. 
If this is so, biological populations are expected to evolve mainly within 
these clusters and consist most of the time of well-fit genotypes with 
fitnesses within some band. The metaphor of ``holey'' adaptive landscapes 
neglects the fitness differences between genotypes in the cluster but these
differences are supposed to exist and should be apparent on a finer scale. 
If one applies a finer resolution, the movement along the cluster will be 
accompanied by slight increases or decreases in fitness. Evolution will 
proceed by fixation of weakly selected alleles which can 
be advantageous, deleterious, over- and underdominant, or apparently 
neutral depending on the specific area of genotype space the population passes
through. Smaller populations will pass faster through the areas of genotype space
corresponding to fixation of slightly deleterious mutations whereas larger
populations will pass faster through the areas corresponding to fixation of (compensatory) slightly advantageous mutations. 
This pattern of molecular evolution, as predicted from the general 
properties of multidimensional adaptive landscapes, is similar to the 
patterns revealed by the methods of experimental molecular biology, which 
form the empirical basis for the nearly neutral theory of molecular 
evolution (Ohta 1992). From general considerations,
one should not expect complete symmetry of ``real'' adaptive landscapes which 
are supposed to have areas varying with respect to the ``width'' and 
concentration of ridges of well-fit genotypes. Sexual populations are expected 
to spend more time in areas of high concentration of well-fit genotypes 
(Peliti and Bastolla 1994). One of the biological manifestations of
this effect will be apparent reduction in the probability of harmful 
mutations, that is, evolution of genetic canalization (cf., Wagner 1996).
The metaphor of holey adaptive landscapes may be useful for thinking
about these and other evolutionary problems.

\begin{center}
Acknowledgments. 
\end{center}

I am grateful to Chris Boake, Marc Camara, Mitch Cruzan, Nick Barton and
G\"{u}nter Wagner for very helpful comments and suggestions. 
Nick Barton and G\"{u}nter Wagner forced me to extend the generality
of the mathematical approximations developed here. Collaboration with
Hai Li and Michael Vose was crucial in developing a computer program used 
to obtain results of individual-based simulations presented in Figures 3 and 4. 
This work was partially supported by grants from Universit\'{e} P. et 
M. Curie and \'{E}cole Normale Sup\'{e}rieur, Paris, and by National 
Institutes of Health grant GM56693.\\

\renewcommand{\thesection}{\Alph{section}}
\setcounter{secnumdepth}{1}
\setcounter{section}{0}
\setcounter{equation}{0} 
\renewcommand{\theequation}{A\arabic{equation}}

\begin{center}
{\bf Appendix}
\end{center}

Effects of mutation, migration and drift on the dynamics of the average
genetic distances within and between subpopulations have been previously
studied thoroughly (e.g., Watterson 1975; Li 1976; Slatkin 1987;
Strobeck 1987). What is left is to add reproductive isolation 
(that is selection) to the model. I will use the deterministic 
framework assuming that the population size $N\rightarrow \infty$.

{\em The distribution of $D_w$ under rare-alleles and linkage 
equilibrium approximation.}
I will use the standard notations ${\bf A_i}$ and ${\bf a_i}$ for alternative
alleles at the $i$-locus ($i=1,\dots ,L$). Let $p_i$ be the frequency of 
allele ${\bf A_i}$ at the $i$-th locus,
$q_i=1-p_i$, and $\psi_{w,i}=2p_iq_i$. Variable $\psi_{w,i}$ can be thought of as the
probability that two randomly chosen individuals (sequences) from the same
subpopulation are different
at the $i$-th locus. 
Let $d_{w,i}=(l_i^{\alpha}-l_i^{\beta})^2$ be the genetic distance at the $i$-th
locus between two randomly chosen individuals $\alpha$ and $\beta$. Note
that $d_{w,i}=1$ with probability $\psi_{w,i}$ and $d_{w,i}=0$ with probability $1-\psi_{w,i}$.
Because $d_{w,i}$ is a Binomial random variable, its generating function is
$\gamma_{d_{w,i}}(s)=\psi_{w,i} s+1-\psi_{w,i}$, which can be approximated as 
$e^{\psi_{w,i}(s-1)}$ if $\psi_{w,i}<<1$ (rare-alleles approximation). Under
approximate linkage equilibrium, the generating function of $d_w=\sum 
d_{w,i}$ is
	\begin{equation}
             \gamma_{d_w}
(s)=\Pi_i e^{\psi_{w,i}(s-1)}=e^{ \sum_i \psi_{w,i} (s-1)}=e^{D_w(s-1)}
	\end{equation}
where $D_w=\sum_i \psi_{w,i}$. This shows that random variable $d_w$ has 
approximately Poisson distribution with parameter $D_w$ and, thus,
        \begin{equation}
            P(d_w=i) = exp( -D_w) \frac{D_w^i}{i!}.
        \end{equation}

{\em Selection within an isolated population}.
Let $w(d)$ be the expected number of fertile and viable offspring that
can be produced by a pair of individuals different in $d$ loci.
The average fitness of the population is
    \[
         \overline{w} = \sum_j w(j)P(d=j).
    \]
The dynamics of the general model of fertility selection and premating
isolation in a haploid population considered here are identical to that
of a symmetric viability selection model for a diploid population with
viabilities $w(d)$ depending on the number of heterozygous loci $d$.
Under approximate linkage equilibrium, changes in allele frequencies
are described by Wright's equation
   \begin{equation}  \label{wright}
        \Delta_s p_i = \frac{p_iq_i}{2} \frac{ \partial \ln \overline{w}}
                                      { \partial p_i}.
   \end{equation}
(Wright 1969). Using the equalities $\partial \ln \overline{w}/ \partial p_i=2(q_i-p_i)
\partial \ln \overline{w}/ \partial \psi_i$ and $D_w=\sum_i \psi_i$, equation
(\ref{wright}) can be rewritten as
   \begin{subequations} \label{delta}
        \begin{equation} 
            \Delta_s p_i = s p_iq_i(p_i-q_i), \label{delta-a}
        \end{equation}
with 
        \begin{equation}
            s=  \frac{ d \ln \overline{w}}{ d D_w}.
        \end{equation}
   \end{subequations}
To describe the dynamics of allele frequencies one needs to know
the mean fitness of the population. 

{\em Truncation selection}.  This is a selection scheme analyzed in detail 
in the main body of the paper. Here
    \begin{equation} \label{fit}
         w(d) = \left\{ \begin{array}{cc}
                        1 & \mbox{for $d\leq K$},\\
                        0 & \mbox{for $d>K$}.
                       \end{array}
                \right.
    \end{equation}
Using the Poisson approximation (A1), the mean fitness is
      \[
         \overline{w}_{threshold} 
                      = \sum_{i=0}^K exp(-D_w)\frac{D_w^i}{i!} =
                         \frac{ \Gamma(K+1,D_w)}{\Gamma(K+1)},
      \]
where the last equality follows from equation (8.352) in Gradshteyn and 
Ryzhik (1994), resulting in $s$ given by equation (\ref{S}). 
To find equation (\ref{single}), one
starts with (\ref{delta-a}) and proceeds using the fact that 
$\Delta \psi_i \approx 2(q_i-p_i)\Delta p_i$
and that $D_w=\sum \psi_i$. 

Other selection schemes can be considered in a similar way,
and some of them result in relatively compact expressions for $\overline{w}$
and $s$. 

{\em Linear selection}. Here
      \begin{equation}
         w(d) = \left\{ \begin{array}{cc}
                        1-ad & \mbox{for $d\leq K$},\\
                        0    & \mbox{for $d>K$}.
                       \end{array}
                \right.
     \end{equation}
The mean fitness is
      \[
         \overline{w}_{linear} = \overline{w}_{threshold}-a \frac{ D_w \Gamma(K,D_w)}{\Gamma(K)}.
      \]

{\em Quadratic selection}. Here
      \begin{equation}
         w(d) = \left\{ \begin{array}{cc}
                        1-ad-bd^2 & \mbox{for $d\leq K$}\\
                        0      & \mbox{for $d>K$}
                       \end{array}
                \right.
     \end{equation}
The mean fitness is
      \begin{eqalignno*}
         \overline{w} =   \overline{w}_{linear} -  b & \left[ D_w(D_w-K) \right. \\ 
               & + \frac{D_w(K+1)\Gamma(K,D_w)}{\Gamma(K)} \\ 
               & \left. - exp(-D_w) \frac{ D_w^{K+2}H(2,K+2,D_w)}{\Gamma(K+2)}\right],
      \end{eqalignno*}
where $H$ is the hypergeometric function  (Gradshteyn and Ryzhik 1994).

{\em Exponential selection}. Here
      \begin{equation}
         w(d) = \left\{ \begin{array}{cc}
                        exp(-ad) & \mbox{for $d\leq K$},\\
                        0    & \mbox{for $d>K$}.
                       \end{array}
                \right.
     \end{equation}
The mean fitness is
      \[
         \overline{w} = exp( -D_w(1-e^{-a})) \frac{ \Gamma(K+1, D_we^{-a})}
                                             {\Gamma(K+1)}.
      \]
I have not explored how assuming these selection schemes
would affect the outcome of the dynamics.

{\em Stochastic transitions in an isolated population}.
Adding mutation results in equation
    \begin{equation} \label{msb}
         \Delta p_i = sp_iq_i(p_i-q_i) + \mu (q_i-p_i),
    \end{equation}
where $\mu$ is the rate of mutation (assumed to be equal for forward and
backward mutations).
Equation (\ref{msb}) is similar to the classical equation describing 
underdominant selection on a single locus in a diploid population. This allows
one to use Lande's results (1979; see also Hedrick 1981; Walsh 1982; Barton and
Rouhani 1987) to find the rate of stochastic
divergence. This rate is twice the expected number, $vN$, of new mutations in
a population times the probability that a given one will be fixed, $U$.
Using the diffusion approximation, $U$ is defined by equations (1a) and
(2) in (Lande 1979). Lande used some approximations to evaluate $U$.
However, the integrals in his equation (1a) can be found exactly resulting
in  
    \begin{equation}   \label{exac}         
      U = \frac{1}{2} 
   \left(1- \frac{ erf \left[ \sqrt{S}(1-\frac{2}{N}) \right]}
                 { erf \left[ \sqrt{S}\right]} 
   \right),
    \end{equation}
where $S=Ns/2$ (Walsh 1983). Expanding in a Taylor series under
the assumption that $1/N<<1$ results in (\ref{isol-b}) which is equivalent
to Lande's (1979) formula. The difference between Lande's approximate
formula and the exact equation (\ref{exac}) is negligible.

{\em The distribution of $D_b$ under rare-alleles and linkage 
equilibrium approximation}.
Let us consider two subpopulations. Let $p_i$ and $P_i$ be the frequencies
of allele ${\bf A_i}$ in the first and second subpopulations, respectively.
The genetic distance $d_{b,i}$ at the $i$-th locus between two randomly
chosen sequences from two different subpopulations is a Binomial variable
taking values $1$ and $0$ with probabilities  $\psi_{b,i}=p_iQ_i+q_iP_i$
and $1-\psi_{b,i}$, respectively ($q_i=1-p_i, Q_i=1-P_i$). I will assume 
that genetic variation within each subpopulation is low so that 
$\psi_{b,i}$ is close to either $0$ or $1$. Let
$\delta_i=d_{b,i}$ if $\psi_{b,i} \approx 0$ and 
$\delta_i=1-d_{b,i}$ if $\psi_{b,i} \approx 1$. The genetic
distance between individuals $\alpha$ and $\beta$ can be represented as
$d_b=k-\sum_1 \delta_i+\sum_2 \delta_i$ where the first sum is over 
$k$ loci at which $\psi_{b,i} \approx 1$ and the second sum is over $L-k$ 
loci at which $\psi_{b,i} \approx 0$. Using the assumption of linkage 
equilibrium, the generating function of $d_b$ becomes
	\begin{eqalignno*}
		\gamma_{d_b}(s)= & E\{ s^{k-\sum_1 \delta_i+\sum_2 \delta_i}\}\\
                 & =s^k \Pi_{i=1}^k
e^{-\phi_i(s-1)} \Pi_{i=k+1}^L e^{\phi_i(s-1)} \\ & = \frac{s^k}{e^{k(s-1)}}
e^{D_b(s-1)},
	\end{eqalignno*}
where $\phi_i$ is the expectation of $\delta_i$ and $D_b$ is the 
expectation of $d_b$. Using the properties of generating functions, 
the distribution of $d_b$  is
    \begin{equation} \label{betw}
         P(d_b=i) = \left\{ \begin{array}{cc}
                        0 & \mbox{if $i<k$},\\
                        \frac{(D_b-k)^{i-k}}{(i-k)!}e^{-(D_b-k)} & \mbox{if $i\geq k$}.
                       \end{array}
                \right.
    \end{equation}
With fitness function (\ref{fit}), the probability that two randomly 
chosen individuals from different subpopulations are not reproductively
isolated is
	\begin{equation}
		\overline{w}_b = \sum_{i=0}^K P(d_b=i)=\frac{ \Gamma(K-k+1,
			D_b-k)}{\Gamma(K-k+1)},
	\end{equation}
if $k\leq G$ and $\overline{w}_b=0$ if $k>K$.

{\em Deterministic dynamics in a subdivided population}. 
With no reproductive isolation and with equal forward and backward
migration rates and equal population sizes, the change in $p_i$ due to 
migration is $\Delta_m p_i = m(P_i-p_i)$. The corresponding change in 
$D_w$ is $\Delta_m D_w = 2m(D_b-D_w)$. 
With reproductive isolation and given that $D_b \geq D_w$, individuals 
migrating 
from other subpopulations will have reduced probability of mating. Let 
$\overline{w}_w$ and $\overline{w}_b$ be the expected numbers of fertile and
viable offspring that can be produced as a result of within and between subpopulations encounters. For simplicity I will omit
the index specifying the locus under consideration.
With equal population sizes and migration rates,
the change in the allele frequency due to migration becomes
  \begin{subequations}
    \begin{equation}
         \Delta_m p = m_e (P-p),
    \end{equation}
where the ``effective'' migration rate is
    \begin{equation}
         m_e = m\ \frac{\overline{w}_b}{\overline{w}_w}.
    \end{equation}
(compare with the models with migration between populations of unequal size
where the effective migration rate is $m$ times the ratio of the population
sizes, e.g. Gavrilets 1996). The corresponding change in $D_w$ is
    \begin{equation}
         \Delta_m D_w = 2m_e (D_b-D_w).
    \end{equation}
  \end{subequations}
Changes $\Delta_m p_i$ can be thought of as changes in
allele frequencies brought about by selection between groups of individuals
(migrants and residents) whereas the first term in the right-hand side
of equation (\ref{msb})  can be thought of as the change
in $p$ brought about by individual selection.

The dynamics of allele frequencies at a specific locus
under the joint action of selection, mutation and migration are described by
   \begin{subequations} \label{det}
   \begin{eqalignno}
      \Delta p & = s pq(p-q)+m_e(P-p)+\mu(q-p),\\
      \Delta P & = \tilde{s} PQ(P-Q)+m_e(p-P)+\mu(Q-P).
   \end{eqalignno}
   \end{subequations}
With $s=\tilde{s}=const$ and $m_e=const$ and $m_e<s/6$, dynamic system
(\ref{det}) has two types of stable equilibria: mutation-selection balance equilibria with $p \approx P,\ pq \approx
\mu/s,\ \psi_w \approx 2\mu/s, \psi_b \approx 2\mu/s$ and migration-selection
equilibria with $p \approx Q,\ pq \approx
\mu/s+m/s,\ \psi_w \approx 2\mu/s +2m/s, \psi_b \approx 1-2\mu/s-2m/s$.
I assume that in the deterministic limit, $k$ out of $L$ loci evolve towards
migration-selection balance equilibria whereas the remaining $L-k$ loci
evolve towards mutation-selection balance equilibria. In the latter $L-k$ loci,
the dynamics of $\psi_w$ and $\psi_b$ are approximated by equations
     \begin{subequations} \label{psi}
       \begin{eqalignno}
          \Delta \psi_w & = - s \psi_w + 2 \mu + 2m_e(\psi_b-\psi_w), \label{psi-a}\\
          \Delta \psi_b & = - s \psi_b + 2 \mu + 2m_e(\psi_w-\psi_b). \label{psi-b}
       \end{eqalignno}
     \end{subequations}
In the former $k$ loci, the dynamics of $\psi_w$ are described as before
by (\ref{psi-a}) whereas the dynamics of $\psi_b$ are approximated by equation
       \begin{equation} \label{psib}
          \Delta \psi_b =  s (1 - \psi_b) - 2 \mu + 2m_e(\psi_w-\psi_b).
       \end{equation}
Selection always reduces $\psi_w$ whereas mutation always increases it (see
equation \ref{psi-a}).
Selection and mutation have the same effects on $\psi_b$ for the loci evolving
toward mutation-selection balance equilibria (see
equation \ref{psi-b}). However, for the loci evolving
toward migration-selection balance equilibria, selection increases $\psi_b$ 
whereas mutation decreases it (see
equation \ref{psib}). Summing up over all loci, one finds equations 
(\ref{migr}) 
of the main text. Equations (\ref{per}) are derived in a similar way
assuming that the allele frequencies in the main population do not change.

{\em Stochastic transitions in a subdivided population}.
In a subdivided population, migration tends to reduce genetic differentiation.
Given that migration is sufficiently strong relative to selection, 
the same allele will be close to fixation in both subpopulations. If, by a
chance, an alternative allele approaches fixation in one of the subpopulations
creating significant differentiation at a given locus, such differentiation
will be quickly lost. The number of loci $k$ at which alternative alleles
are close to fixation in different subpopulations will be close to zero on
average. However, if migration is relatively weak, then the differentiation
created by random genetic drift will not be lost quickly and actually can even
accumulate. Let us consider a locus at which initially the same allele is
close to fixation in both subpopulations
(that is both $p \approx 0$ and $P \approx 0$). Neglecting
the changes in $P$, the deterministic change in $p$ due to selection and
migration is approximately
    \begin{equation}
             \Delta p = spq(p-q)-m_ep.
    \end{equation}
Lande (1979; see also Barton and Rouhani 1987) has shown that the rate 
at which allele {\bf A} becomes close to fixation in 
the first subpopulation while its frequency is about zero in the second
population is approximately $2^{-Nm_e}$ times the rate of
fixation in the absence of immigration. Assuming that alleles {\bf A}
are brought about by mutation at rate $\mu$ and summing up over $L-k$ 
loci, one
finds the first term in the right-hand side of equation (\ref{k}). Once
alternative alleles are close to fixation in different subpopulations,
random drift can remove genetic differentiation. Let us consider a locus at 
which initially $p \approx 0$ but $P \approx 1$. Neglecting
the changes in $P$, the deterministic change in $p$ due to selection and
immigration is approximately
    \begin{equation}
             \Delta p = spq(p-q)-m_e(1-p).
    \end{equation}
Using Barton and Rouhani's (1987) method one finds that the rate
at which allele {\bf A} becomes close to fixation in 
both subpopulations is approximately $(e/2)^{Nm_e}$ times the rate of
fixation in the absence of immigration. Assuming that alleles {\bf A}
are brought about by migration at rate $m_e$ and summing up over $k$ loci, one
finds the second term in the right-hand side of equation (\ref{k}).

{\em Stochastic divergence with local adaptation}. Let us assume that the 
allele under consideration is favorable in a given environment with selective advantage $s_{LA}$.
The change in this allele frequency as defined by the joint action of selection
induced by reproductive isolation and selection for local adaptation is
    \begin{equation} \label{RIvsLA}
        \Delta_s p = spq(p-q)+ s_{LA}pq.
    \end{equation}
This equation is identical to the one describing meiotic drive
in the Appendix of Walsh (1982). Following Walsh, the fixation probability is
   \begin{equation}   \label{exact}         
      U = \frac{ erf \left[ \sqrt{S}(1-\alpha)\right]
                -erf \left[ \sqrt{S}(1-\alpha-\frac{2}{N}) \right]}
               { erf \left[ \sqrt{S}(1-\alpha)\right] 
                +erf \left[ \sqrt{S}(1+\alpha) \right] },
    \end{equation}
where $S=Ns/2$ and $\alpha=s_{LA}/s$. Expanding the numerator in a Taylor 
series under the assumption
that $1/N<<1$ and multiplying the results by the expected number of 
mutants, $vN$, results in the relative rate of fixation given by equation
(\ref{walsh}).

{\em Genetic distance (\ref{gen_dist}) and some other models}.
Genetic distance (1) is recovered by assuming that $G$ is an identity
matrix. Assuming that $G$ is a diagonal matrix with non-equal diagonal 
elements is a simple way to introduce non-equivalence of loci. The case 
when the probability of mating depends on the difference in a quantitative 
trait can be treated within the same framework. Let $c_i$ be the contribution 
of the $i$-th locus to a quantitative trait $z$. 
Neglecting microenvironmental effects, the
trait values for individuals $\alpha$ and $\beta$ are $x^{\alpha}=\sum 
c_il_i^{\alpha}$ and $z^{\beta}=\sum c_il_i^{\beta}$, respectively. The square
of the difference of $z^{\alpha}$ and $z^{\beta}$ is recovered from equation
({\ref{gen_dist}) by assuming that $G_{ij}=c_ic_j$. A common way to model
sexual selection is to assume that the probability of mating between a male 
and a female depends on the difference in a female phenotypic trait, $z_f$, 
and a male phenotypic trait, $z_m$, which are controlled by two different 
sets of loci (e.g., Lande 1981; Kirkpatrick 1982; Nei at al. 1983; Wu 1985;
Turner and Burrows 1995). Let $z_m=\sum  c^m_il_i$ and $z_f=\sum c^f_il_i$ 
where the sums are taken over the corresponding sets of loci. The 
value $(z_m-z_f)^2$ is recovered from (\ref{gen_dist}) by assuming that 
matrix $G$ has a block form 
   \[
       G= \left( \begin{array}{cc} 0 & G^s \\ G^s & 0 \end{array} \right).
   \]
The diagonal $L_m\times L_m$ and $L_f\times L_f$ zero matrices correspond to 
the interactions within the set of $L_m$ genes controlling the male trait 
and within the set of $L_f$ genes controlling the female trait 
($L=L_f+L_m$), and matrix $G_s$ describing the interactions between 
the two sets of genes has elements $G^s_{ij}=c_i^f c_j^m$. 

\newpage
{\bf Literature Cited}

  Arnold, M.L. 1997.  Natural Hybridization and Evolution. Oxford
University Press, Oxford. 

  Ayala, F.J., M.L. Tracey, D. Hedgecock, and R.C. Richmond. 1974.
Genetic differentiation during the speciation process in {\em Drosophila}.
Evolution 28: 576-592.

Barton, N.H. 1986. The maintenance of polygenic variation through a balance
between mutation and stabilizing selection. Genet. Res. 47: 209-216.

  Barton, N.H. 1989. Founder effect speciation. In  Speciation and its 
consequences (Otte D. and Endler J.A., eds) Sunderland, Massachusetts. 
pp. 229-256.

Barton, N.H., and B.O.Bengtsson. 1986. The barrier to genetic exchange
between hybridizing populations. Heredity 56: 357-376.
 
  Barton, N.H., and B. Charlesworth. 1984. Genetic revolutions,
 founder effects, and speciation. Ann. Rev. Ecol. Syst.
 15: 133-164.
 
  \label{BaRo87} Barton, N.H., and S. Rouhani. 1987. The frequency of shifts
between alternative equilibria. J. Theor. Biol. 125: 397-418.

Barton, N.H. and M. Turelli. 1987. Adaptive landscapes, genetic distance
and the evolution of quantitative characters. Genet. Res. 49: 157-173.

Bengtsson, B.O. 1985. The flow of genes through a genetic barrier. 
In Evolution Essays in Honor of John Maynard Smith (Greenwood J.J., 
P.H.Harvey, and M.Slatkin, eds). Cambridge University Press, Cambridge. pp.31-42. 

Bengtsson B.O. and Christiansen, F.B. 1983. A two-locus mutation 
selection model and some of its evolutionary implications. Theor. Popul.
 Biol. 24: 59-77.

  Birky, C.W., and J.B. Walsh. 1988. Effects of linkage on rates of
molecular evolution. Proc. Natl. Acad. Sci. USA 85: 6414-6418.

\label{Bull94} Bullini, L. 1994. Origin and evolution of 
animal hybrid species. Trends Ecol.Evol. 9: 422-426.

B\"{u}rger, R., G.P. Wagner and F. Stettinger. 1989. How much heritable variation can be maintained in finite population by mutation-selection
balance? Evolution 43: 1748-1766.

  \label{Clar97} Claridge, M.F., H.A. Dawah, and M.R. Wilson  (eds). 1997.
Species. The Units of Biodiversity. Chapman and Hall, London.

  \label{Coyn92} Coyne, J.A. 1992. Genetics and speciation. 
Nature 355: 511-515. 

  \label{Coyn97} Coyne, J.A., N.H. Barton, and M. Turelli. 1997. 
A critique of Sewall Wright's shifting balance theory of evolution.
 Evolution 51: 643-671.

  del Solar, E. 1966. Sexual isolation caused by selection for
positive and negative phototaxis and geotaxis in {\em Drosophila 
pseudoobscura}. Proc Natl. Acad. Sci. USA 56: 484-487.

  Derrida, B., and L. Peliti. 1991. Evolution in a flat fitness landscape.
Bull. Math. Biol. 53: 355-382.

  \label{Dobz37} Dobzhansky, T.H. 1937.  Genetics and the Origin of 
Species. Columbia University Press. 

  Dodd, D.M.B. 1989. Reproductive isolation as a consequence of
adaptive divergence in {\em Drosophila pseudoobscura}. Evolution 43: 
1308-1311.

  Garcia-Ramos, G., and M. Kirkpatrick. 1997. Genetic models of
adaptation and gene flow in peripheral populations. Evolution 51: 21-28.

  \label{Gavr96} Gavrilets, S. 1996. On phase three of the shifting-balance 
theory. Evolution 50: 1034-1041.

  \label{Gavr97a} Gavrilets, S. 1997a. Evolution and speciation on holey 
adaptive landscapes. Trends Ecol. Evol. 12: 307-312.

  \label{Gavr97b} Gavrilets, S. 1997b. Hybrid zones with Dobzhansky-type
epistatic selection. Evolution 51: 1027-1035.

 Gavrilets, S., and G. de Jong. 1993. Pleiotropic models of polygenic
variation, stabilizing selection and epistasis. Genetics 134: 609-625.

  \label{GaHa96} Gavrilets, S., and A. Hastings. 1996.
Founder effect speciation: a theoretical reassessment. Am. Nat. 147: 
466-491 1996. 

  \label{GaGr97} Gavrilets, S., and J. Gravner. 1997.    Percolation 
on the fitness hypercube and the evolution of reproductive isolation. 
J. Theor. Biol. 184: 51-64.

  Gavrilets, S., and C.R.B. Boake. 1998. On the evolution of premating 
isolation after a founder event. Amer. Natur. 000: 000-000.

  Gavrilets, S., H. Li, M.D. Vose. 1998. Rapid speciation on holey
adaptive landscapes. Proc. R. Soc. Lond. B, 000: 000-000.

  Gillespie, J.H. 1991. The causes of molecular evolution. Oxford 
University Press, New York.

  Givnish, T.J., and K.J. Sytsma (eds). 1997. Molecular evolution and 
adaptive radiation. Cambridge University Press, Cambridge.

  \label{GrRy94} Gradshteyn, I.S., and I.M. Ryzhik. 1994. Tables of
Integrals, Series, and Products. Fifth Edition. Academic Press, San Diego.

  Haldane, J.B.S. 1927. A mathematical theory of natural and artificial
selection. V. Selection and mutation. Camb. Phylos. Soc. Proc. 23: 838-844.

  Hedrick, P.W. 1981. The establishment of chromosomal variants. Evolution
35: 322-332.

  Higgs, P.G., and B. Derrida. 1991. Stochastic model for species
formation in evolving populations. J. Phys. A: Math. Gen. 24: L985-L991. 

   \label{Higg92} Higgs, P.G., and B. Derrida. 1992.
Genetic distance and species
formation in evolving populations. J. Mol. Evol. 35: 454-465.

  \label{phyl} Hoelzer, G.A., and D.J. Melnick. 1994. Patterns of 
speciation and
limits to phylogenetic resolution. Trends in Ecology and Evolution
9: 104-107.

  \label{John96}Johnson, T.C. et al. 1996. Late Pleistocene desiccation of 
Lake Victoria and rapid evolution of
cichlid fishes. Science 273: 1091-1093.

  \label{KaMc72} Karlin S., and J. McGregor. 1972. Application of method of small parameters in multi-niche population genetics models. Theor. Popul. Biol. 3: 180-209.  

  Kilias, G., S.N. Alahiotis, and M. Pelicanos. 1980. A multifactorial
genetic investigation of speciation theory using {\em Drosophila melanogaster}.
Evolution 34: 730-737.

  Kirkpatrick, M. 1982. Sexual selection and the evolution of female choice.
Evolution 36: 1-12.

Kondrashov, A.S., and Mina. 1986.  Sympatric speciation: when is it
possible? Biol. J. Linn. Soc. 27: 201-223.

  Lande R. 1976. Natural selection and random genetic drift in
phenotypic evolution. Evolution 30: 314-334.

  Lande, R. 1979. Effective deme sizes during long-term evolution estimated
from rates of chromosomal rearrangements. Evolution 33: 234-251.

  Lande, R. 1981. Models of speciation by sexual selection on polygenic
traits. Proc. Natl. Acad. Sci. USA 78: 3721-3725.

  \label{Li76} Li, W.-H. 1976. Distribution of nucleotide differences between
two randomly chosen cistrons in a subdivided population: the finite island
model. Theor. Popul. Biol. 10: 303-308.

  Li, W.-H. 1997. Molecular Evolution. Sinauer Associates: Sunderland,
Massachusetts.

  \label{Mall95} Mallet, J. 1995. A species definition for the 
modern synthesis. Trends Ecol. Evol. 10: 294-299.

  \label{Manz94} Manzo, F., and L. Peliti. 1994.
 Geographic speciation in the
Derrida-Higgs model of species formation. J. Phys. A: Math. Gen. 27:
7079-7086.

  Mani, G.S., and B.C.C. Clarke. 1990. Mutational order: a major
stochastic precess in evolution. Proc. R. Soc. Lond. B 240: 29-37.

\label{Mayn70} Maynard Smith, J. 1970. Natural selection and 
the concept of a protein space.  Nature 225: 563-564.

Maynard Smith J. 1983. The genetics of stasis and punctuation. Ann. Rev.
Genet. 17: 11-25.

  Mayr, E. 1963. Animal species and evolution. Belknap Press, Harvard,
Cambridge, MA.

  Mayr, E. 1982a. The growth of biological thought. Harvard University
Press, Cambridge, MA.

  Mayr, E. 1982b. Speciation and macroevolution. Evolution 36:
1119-1132.

  \label{McCu96} McCune, A.R. 1996. Biogeographic and stratigraphic evidence 
for rapid speciation in semionotid fishes. Paleobiology 22:
34-48.

    McCune, A.R. 1997. How fast is speciation? Molecular, geological, and
phylogenetic evidence from adaptive radiations of fishes.
Pp 585-610 {\em in} Givnish, T.J., and K.J. Sytsma (eds). Molecular 
evolution and adaptive radiation. Cambridge University Press, Cambridge.

Nei, M. 1976. Mathematical models of speciation and genetic distance.
pp.723-768 in S. Karlin and E. Nevo (eds.) Population Genetics and Ecology.
Academic Press: NY.

  \label{Nei83} Nei, M., T.Maruyma, and C-I. Wu. 1983. Models of evolution of
reproductive isolation.  Genetics  103: 557-579. 

  Ohta, T. 1992. The nearly neutral theory of molecular evolution.
Ann. Rev. Ecol. Syst. 23: 263-286.

  \label{Orr95} Orr, H.A. 1995.  The population genetics of 
speciation: the evolution of hybrid incompatibilities. Genetics 
139: 1803-1813.
 
Orr, H.A. 1997.  Dobzhansky, Bateson, and the genetics of speciation.
Genetics 144: 1331-1335.

  \label{Orr96} Orr, H.A., and L.H. Orr. 1996. Waiting for speciation: the effect
of population subdivision on the waiting time to speciation. Evolution
50: 1742-1749.

\label{Prov86} Provine, W.B. 1986. Sewall Wright and Evolutionary Biology, 
The University of Chicago Press, Chicago.

  \label{PeBa94} Peliti, L., and U. Bastolla. 1994. Collective 
adaptation in a statistical model of an evolving population. C. R. Acad.
Sci. Paris, Sciences de la vie 317: 371-374.

  \label{Reid97} Reidys, C.M., P.F. Stadler, and P. Schuster. 1997. 
Generic
properties of combinatory maps: neutral networks of RNA secondary
structures. Bull. Math. Biol. 59: 339-397.

  Reinthal, P.N., and A. Meyer. 1997. Molecular phylogenetic tests of
speciation models in Lake Malawi cichlid fishes. Pp 375-390 {\em in} 
Givnish, T.J., and K.J. Sytsma (eds). Molecular evolution and 
adaptive radiation. Cambridge University Press, Cambridge.

  Rice, W.R. 1984. Disruptive selection on habitat preferences and
the evolution of reproductive isolation: a simulation study. Evolution 38:
1251-1260.
 
  Rice, W.R., and E.E. Hostert. 1993. Laboratory experiments on
speciation: what have we learned in 40 years? Evolution 47: 1637-1653.

  Rice, W.R., and G.B. Salt. 1988. Speciation via disruptive selection
on habitat preference: experimental evidence. Amer. Natur. 131: 911-917.

Rieseberg, L.H. 1995. The role of hybridization 
in evolution: old wine in new skins. Am. J. Bot. 82: 944-953. 

Ringo, J., D. Wood, R. Rockwell, and H. Dowse. 1985. An experiment testing
two hypothesis of speciation. Amer. Natur. 126: 642-651.

  \label{Schlu96} Schluter, D. 1996. Ecological causes of adaptive radiation. Amer. Natur. 148: S40-S64.

  \label{ScMc92} Schluter, D., and J.D. McPhail. 1992. Ecological displacement
and speciation in sticklebacks.  Amer. Natur. 140: 85-108.

  \label{Slat87} Slatkin, M. 1987. The average number of sites segregating DNA
sequences drawn from a subdivided population. Theor. Popul. Biol.
32: 42-49.

  \label{Stro87} Strobeck, C. 1987. Average number of nucleotide differences 
in a
sample from a single subpopulation: a test for population subdivision. 
Genetics 117: 149-153.

  Templeton, A.R. 1996. Experimental evidence for the genetic
 transilience model of speciation. Evolution 50: 909-915.

  Turner, G.F., and M.T. Burrows. 1995. A model of sympatric speciation
by sexual selection. Proc. R. Soc. Lond. B 260: 287-292.

  Wagner A. 1996. Does evolutionary plasticity evolve? Evolution 50:
1008-1023.

  \label{Wagn94} Wagner A., G.P. Wagner, and P. Similion. 1994. 
Epistasis can facilitate the evolution of reproductive isolation by 
peak shifts - a two-locus two-allele model. Genetics 138: 533-545.

  Walsh, J.B. 1982. Rate of accumulation of reproductive isolation by
chromosome rearrangements. Amer. Natur. 120: 510-532.

  \label{Watt75} Watterson, G.A. 1975. On the number of segregating sites 
in genetic models without recombination. Theor. Popul. Biol. 7:
256-276.

  \label{Whit95} Whitlock, M.C. {\em et al.} 1995.
Multiple fitness peaks and epistasis. Annu. Rev. Ecol. Syst. 26:
601-629.

  Wiley, E.O. 1988. Vicariance biogeography. Ann. Rev. Ecol. Syst. 19:
513-542.

Wills, C. J. 1977. A mechanism for rapid allopatric speciation. Amer. Natur.
111: 603-605.

  \label{Wrig31} Wright, S. 1931. Evolution in Mendelian 
populations. Genetics 16: 97-159.

  \label{Wrig32} Wright, S. 1932. The roles of mutation, inbreeding,
crossbreeding and selection in evolution.  Proc. Sixth
Int. Congr. Genet. 1: 356-366.

  Wright, S. 1969. Evolution and the genetics of populations. Vol.2.
The theory of gene frequencies. University of Chicago Press, Chicago.

  Wright, S. 1982. The shifting balance theory and macroevolution.
Ann. Rev. Genet. 16: 1-19.

  Wu, C-I. 1985. A stochastic simulation study of speciation by
sexual selection. Evolution 39: 66-82.

 Wu, C-I., and M. F. Palopoli. 1994. Genetics of postmating reproductive
isolation in animals. Ann. Rev. Genetics 28: 283-308.

Yampolsky, L.Y., R.M. Kamaltinov, D. Ebert, and others. 1994. Variation
of allozyme loci in endemic gammarids of Lake Baikal. Biol. J. Linn. Soc.
53: 309-323.

\end{document}